\documentclass[pra,showpacs,twocolumn,aps,superscriptaddress,a4paper,longbibliography]{revtex4-1}
\usepackage{dcolumn,amssymb,amsmath,amsfonts,graphicx,latexsym,color}
\usepackage[utf8]{inputenc}
\usepackage[T1]{fontenc}
\usepackage{epstopdf}

\usepackage{ulem}
\usepackage{epstopdf}
\usepackage{subfigure}
\usepackage{float}
\usepackage{mathrsfs}
\usepackage{bbm}
\usepackage{psfrag}

\usepackage{verbatim}
\usepackage{multirow}
\usepackage{diagbox}
\usepackage[colorlinks,citecolor=blue]{hyperref}

\usepackage{CJK}
\begin{document}
\begin{CJK*}{GBK}{song}

\title{Kinked linear response from non-Hermitian cold-atom pumping}
\author{Fang Qin}\email{qinfang@nus.edu.sg}
\thanks{These authors contributed equally to this work.}
\affiliation{Department of Physics, National University of Singapore, Singapore 117551, Singapore}
\author{Ruizhe Shen}\email{ruizhe20@u.nus.edu}
\thanks{These authors contributed equally to this work.}
\affiliation{Department of Physics, National University of Singapore, Singapore 117551, Singapore}
\author{Linhu Li}\email{lilh56@mail.sysu.edu.cn}
\affiliation{Guangdong Provincial Key Laboratory of Quantum Metrology and Sensing $\&$ School of Physics and Astronomy, Sun Yat-Sen University (Zhuhai Campus), Zhuhai 519082, China}
\author{Ching Hua Lee}
\email{phylch@nus.edu.sg}
\affiliation{Department of Physics, National University of Singapore, Singapore 117551, Singapore}
\date{\today}

\begin{abstract}
It is well known that non-Hermitian, non-reciprocal systems may harbor exponentially localized skin modes. However, in this work, we find that, generically, non-Hermiticity gives rise to abrupt and prominent kinks in the semi-classical wave packet trajectories of quantum gases, despite the absence of sudden physical impulses.  This physically stems from a hitherto underappreciated intrinsic non-locality from non-Hermitian pumping, even if all physical couplings are local, thereby resulting in enigmatic singularities in the band structure that lead to discontinuous band geometry and Berry curvature. Specifically, we focus on the realization of the kinked response in an ultracold atomic setup. For a concrete experimental demonstration, we propose an ultracold atomic setup in a two-dimensional optical lattice with laser-induced loss such that response kinks can be observed without fine-tuning in the physical atomic cloud dynamics. Our results showcase unique non-monotonic behavior from non-Hermitian pumping beyond the non-Hermitian skin effect and suggest new avenues for investigating non-Hermitian dynamics on ultracold atomic platforms.
\end{abstract}

\maketitle
\end{CJK*}

\section{Introduction}\label{1}

Ultracold atomic platform simulations for novel condensed-matter phenomena have gained significant attention recently~\cite{li2019observation,lapp2019engineering,ren2022chiral,liang2022dynamic,gou2020tunable}. Our proposed cold-atom setup showcases the potential of synthetic spin-orbit coupling~\cite{lin2011spin,huang2016experimental,wu2016realization,wang2021realization}, laser-induced loss~\cite{ren2022chiral}, and optical lattice potential~\cite{atala2013direct,goldman2016topological,yang2020cooling} as powerful tools for quantum control and simulations. This combination of tools opens up new avenues for investigating novel phenomena, such as response kinks, and exploring exotic quantum phases in ultracold atomic systems.

Far beyond the well-known non-Hermitian skin effect from non-reciprocal hopping, in this work, we find that we should also expect very prominent and abrupt jumps in wave packet trajectories, even though the system does not contain any sudden impulses. Specifically, we focus on the realization of the kinked response in an ultracold atomic setup. Response kinks are physically interesting because they signal abrupt, unexpected changes in a system's behavior. This abruptness often leads to emergent phenomena, critical insights, and practical applications in fields like materials science and engineering. These kinks are surprising because they correspond to divergently large semi-classical impulses, even though no such physical impulses exist. We trace their existence to an emergent form of non-locality (even though the physical system is completely local) caused by directed non-Hermitian pumping, which is already known for producing boundary-localized skin modes~\cite{yao2018edge,lee2019anatomy,helbig2020generalized,okuma2020topological,jiang2022dimensional,zhang2022universal,yokomizo2020non}. Mathematically, the emergent non-locality can be encoded in the non-analyticity of the generalized Brillouin zone (GBZ)~\cite{qin2023universal}, where non-Hermitian pumping causes the effective lattice momentum $k\rightarrow k+i\kappa(k)$ to be generically complex-deformed by an envelope factor $\kappa(k)$ that contains cusp singularities in all but the simplest models. As such, all quantities that contain momentum-space gradients, such as the band metric and Berry curvature, become discontinuous. Thus, kinks appear in their associated linear responses, with the very large ``virtual'' impulses encoding the emergent non-locality.

A caveat is that these response kinks tacitly hinge on the notion of band occupancy and thus cannot appear in metamaterial platforms such as non-Hermitian photonic lattices~\cite{regensburger2012parity,feng2017non,midya2018non} and topolectrical circuits~\cite{helbig2020generalized,hofmann2020reciprocal,ezawa2019electric,liu2021non,shang2022experimental,zhang2023electrical,zhu2023higher,hohmann2023observation}, whose great success are restricted to single-particle phenomena. As such, building on rapid parallel developments in ultracold atomic systems~\cite{li2019observation,lapp2019engineering,ren2022chiral,liang2022dynamic,gou2020tunable}, we formulate a detailed experimental proposal for observing these kinks in the trajectories of atomic clouds.

The paper is organized as follows: In Section~\ref{2}, the general formalism for the non-analytic complex momentum deformation from non-Hermitian pumping is presented. In Section~\ref{3}, we describe how non-analyticities in the bulk momentum can give rise to jumps in the band geometry and topology. In Section~\ref{4}, we discuss response kinks from discontinuous Berry curvature. In Section~\ref{5}, we propose a cold-atom setup to detect kinked responses. Finally, we provide a discussion in Section~\ref{6}.

\section{Non-analytic complex momentum deformation from non-Hermitian pumping}\label{2}

We first review how non-Hermiticity can deform the effective lattice momentum non-analytically into the complex plane. A generic two-dimensional (2D) non-interacting lattice Hamiltonian can be written as
\begin{equation}
H=\sum_{x,y}\left(\sum_{ji;\alpha\beta} t^{\alpha\beta}_{ji}c^{\alpha\dagger}_{j+x,i+y} c^{\beta}_{x,y}\right),
\end{equation}
where $x,y$ and $\alpha,\beta$ respectively index the unit-cell position and sublattices. The hopping amplitude from unit cell $(x,y)$ and sublattice $\beta$ to unit cell $(x+j,y+i)$ and sublattice $\alpha$ is given by $t^{\alpha\beta}_{ji}$. Since a realistic lattice is finite and bounded, we shall consider open boundary conditions (OBCs) with widths $L_x$ and $L_y$, such that $t^{\alpha\beta}_{ji}=0$ whenever $(x,y)$ or $\hspace{-.5mm}(x+j,y+i)\hspace{-.5mm}$ is not within $[1,L_x]\times [1,L_y]$. In the presence of asymmetric couplings ($|t^{\alpha\beta}_{ji}|\neq |t^{\beta\alpha}_{ij}|$), states are amplified in the direction of stronger hopping, leading to the net effect of non-Hermitian pumping, where all states eventually accumulate against the boundaries~\cite{yao2018edge,lee2019anatomy,yokomizo2019non,yang2020non,okuma2020topological,qin2023non,jiang2022dimensional}.

To illustrate how complex-deformed momentum arises, we consider the square lattice in two dimensions for simplicity, such that non-Hermitian pumping in either dimension can be separately treated. For a one-dimensional (1D) subchain $H_\text{1D}(y)= \sum_{x}\sum_{j;\alpha\beta} t^{\alpha\beta}_{j}c^{\alpha\dagger}_{j+x,y}c^{\beta}_{x,y}$, the periodic boundary condition (PBC) spectrum is simply given by the eigenvalues of the matrix symbol $H_\text{1D}(z)$, where
\begin{equation}
H_\text{1D}^{\alpha\beta}(z)=\sum_{j}t^{\alpha\beta}_{j0}z^{j},
\end{equation}
and $z=e^{ik_{x}}$, $k_{x}\in\{2\pi /L_{x},4\pi/L_{x},\cdots\}$. However, under OBCs, the accumulated boundary ``skin'' states drastically break translation invariance, and the correct bulk description can only be obtained by complex deforming $k_{x}\rightarrow\bar k_{x}=k_{x}+i\kappa(k_{x})$ with an appropriate $\kappa(k_{x})$ such that approximate translation invariance (i.e., ``bulk-boundary correspondence''~\cite{yokomizo2019non,yokomizo2020non,yang2020non}) is restored~\cite{kunst2018biorthogonal,xiong2018does,yang2020non}. The deformation $\kappa(k_{x})=-\ln|z|$ is determined from the requirement that all OBC wavefunctions vanish at both ends of $[1,L_{x}]$~\cite{yao2018edge,lee2019anatomy,okuma2020topological,jiang2022dimensional}. This is equivalent to the condition in the dispersion equation
\begin{equation}
\text{Det}\left[H_\text{1D}(z)-E_\text{OBC}\,\mathbb{I}\right]=0,\label{eq:Det}
\end{equation}
which possesses doubly degenerate solutions $z$, $z'$ with $|z|=|z'|$ at OBC eigenenergies $E_\text{OBC}$. This is because the smallest solution branch of $\kappa(k_{x})=-\ln|z|$ controls the dominant decay rate of eigensolutions $\psi_{k_{x}}$, viz. $|\psi_{k_{x}}(x)|\sim |e^{i\bar k_{x}x}|= e^{-\kappa(k_x)x}$, and a superposition of two eigensolutions that decay at identical rates is needed to satisfy OBCs at both left and right boundaries.

\begin{figure}
\centering
\includegraphics[width=1\linewidth]{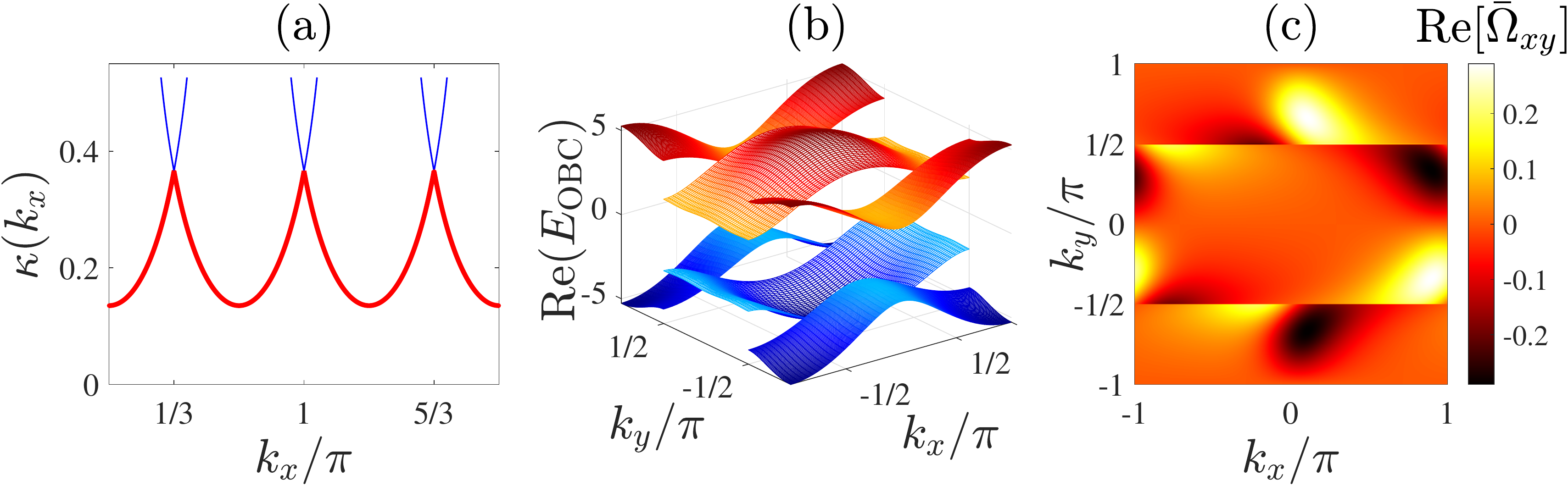}
\caption{\textbf{Discontinuous bands and their Berry curvature from complex momentum singularities.} (a) For our illustrative Hamiltonian $H_\text{Toy}$ [Eq.~\eqref{eq:H1D}] with $\alpha=1$, $\beta=2$, $A=3B$, the effective momentum $\bar k_{x}$ [Eq.~\eqref{eq:H1Dkappa}] acquires an imaginary contribution $\text{Im}[\bar k_x]=\kappa(k_x)$ (red) that is non-analytic at $k_{x}=\pi/3,\pi, 5\pi/3$, wherever different solution branches (blue) intersect. In (b), likewise, the real part of the energy bands (red for the upper band and blue for the lower band) of our cold-atom Hamiltonian $\bar{\cal H}$ defined by Eq.~\eqref{eq:d} exhibits discontinuities at $k_{y}=\pm0.5\pi$ under $\hat{x}$-OBCs and $\hat{y}$-PBCs. (c) The Berry curvature Re$(\bar\Omega_{xy})$ of $\bar{\cal H}$ also exhibits discontinuities at the same locations $k_{y}=\pm0.5\pi$, even though it still integrates to an integer multiple of $2\pi$. Parameters for (b) and (c) are $\phi_{0}=\pi/4$,  $t_{x}=1$, $t_{y}=t_{x}$, $\Omega_{r}=3t_{x}$, $\Omega_{0}=t_{x}$, and $\gamma=t_{x}$. Without loss of generality, these parameters are in arbitrary units (arb. units).}\label{fig:E_model_phy}
\end{figure}

To explicitly see that such $\kappa(k_{x})$ solutions generically possess cusps, we examine the simplest class of asymmetric hopping models with two different nonzero hopping amplitudes $A$, $B$, one $\alpha$ site towards the left and the other $\beta$ site towards the right:
\begin{equation}
H_\text{Toy}(z) = Az^\alpha+\frac{B}{z^\beta},\label{eq:H1D}
\end{equation}
where $z=e^{ik_{x}}$ under PBCs, and $z=e^{i\bar k_{x}}=e^{-\kappa(k_x)}e^{ik_x}$ under OBCs. Since the unit cell is trivial, Eq.~\eqref{eq:Det} reduces to $E_\text{OBC}=Az^\alpha+Bz^{-\beta}$, which is invariant under $z\rightarrow ze^{2\pi i m/(\alpha+\beta)}$, $E_\text{OBC}\rightarrow E_\text{OBC}e^{2\pi i m\alpha/(\alpha+\beta)}$ with $m\in\mathbb{Z}$. As such, $z=z(k_{x})$ and hence $\kappa(k_{x})$ must have a period of $2\pi/(\alpha+\beta)$, i.e., $\kappa\left(k_{x}+\frac{2\pi m}{\alpha+\beta}\right)=\kappa(k_{x})$ with $k_{x}\in [0,2\pi/(\alpha+\beta)]$, $m$ labeling the branch. It can be verified by direct substitution that with the ansatz [Fig.~\ref{fig:E_model_phy}(a)]~\cite{jiang2022dimensional}
\begin{equation}
\kappa(k_{x})=\frac1{\alpha+\beta}\ln\left(\frac{A\sin(\alpha k_{x})}{B\sin(\beta k_{x})}\right),\label{eq:H1Dkappa}
\end{equation}
the red branch with the smallest value of $\kappa(k_{x})$, and hence the slowest spatial decay gives $E_\text{OBC}(k_{x}) =
e^{\frac{2\pi im\alpha}{\alpha+\beta}}\sin[(\alpha+\beta)k_{x}]\times\sqrt[(\alpha+\beta)]{A^\beta B^\alpha /[\sin^{\alpha}(\alpha k_{x})\,\sin^{\beta}(\beta k_{x})]}$ (Appendix \ref{Supp_0}), which collapses into a line segment in the complex energy plane. This implies that any $E_\text{OBC}$ must be visited by two $k_{x}$ values, hence satisfying the requirement of double degeneracy.

Importantly, $\kappa(k_{x})$ as given by Eq.~\eqref{eq:H1Dkappa} generically exhibits kinks where different branches of the smallest $\kappa(k_{x})$ join (except when $\alpha=\beta=1$), since that is where the phase of $E_\text{OBC}$ jumps discontinuously~\cite{li2020critical}. In generic models, there would be multiple solution branches of $\kappa(k_{x})$, leading to inevitable jumps in the definition of complex momentum. Physically, these singularities arise from the competition between the non-local pumpings from asymmetric couplings of different ranges.

\section{Discontinuous Berry curvature}\label{3}

We next discuss how non-analyticities in the bulk momentum description can give rise to jumps in the band geometry and topology. For a square lattice, the effective 2D deformed momentum is $\bar{\bold k}=(k_{x}+i\kappa_{x}(\bold k),k_{y}+i\kappa_{y}(\bold k))$, with $\kappa_{x}$ and $\kappa_{y}$ in general dependent on both momentum components $\bold k = (k_{x},k_{y})$. Consider a two-band 2D Hamiltonian
\begin{equation}
{\cal H}(\bold k)={\cal H}(k_{x},k_{y}) = d_{x}\sigma_{x} + d_{y}\sigma_{y} + d_{z}\sigma_{z}+d_{0}\mathbb{I}, \label{eq:H}
\end{equation}
with the Pauli matrices $\sigma_{x,y,z}$ in pseudospin space. Under OBCs in both $x,y$ directions, the complex-deformed biorthogonal Berry curvature, computed with the occupied left and right momentum eigenstates $\langle\psi_{L}^{}|$ and $|\psi_{R}^{}\rangle$, is
\begin{eqnarray}
\bar{\Omega}_{xy}(\bold k)&\!=\!&\text{Im}\!\left[\langle\partial_{k_x}\psi_{L}^{}(\bar{\bold k})|\partial_{k_y}\psi_{R}^{}(\bar{\bold k})\rangle\right] \nonumber\\
&\!=\!&-\frac{i}{2}\!\left[(\partial_{k_x}D)(\partial_{k_y}{\rm In}R) \!-\! (\partial_{k_y}D)(\partial_{k_x}{\rm In}R) \right]\!,\label{eq:br}
\end{eqnarray}
with $D=d_{z}(\bar{\bold k})/\sqrt{d_{x}^{2}(\bar{\bold k})+d_{y}^{2}(\bar{\bold k})+d_{z}^{2}(\bar{\bold k})}$ and $R=\sqrt{[d_{x}(\bar{\bold k}) - id_{y}(\bar{\bold k})]/[d_{x}(\bar{\bold k}) + id_{y}(\bar{\bold k})]}$. Do note that while the eigenstates $\langle\psi_{L}|$ and $|\psi_{R}\rangle$ are evaluated at complex-deformed momenta $\bar{\bold k}$, their derivatives $\partial_{k_x},\partial_{k_y}$ are taken with respect to the {\it physical} momenta $\bold k$, different from existing definitions~
\cite{shen2018topological,liang2013topological,yokomizo2019non,lieu2018topological}. This is because while $\bar{\bold k}$ compensates for the broken translation invariance from boundary-accumulated pumped states, the derivatives $\partial_{k_x},\partial_{k_y}$ must be taken with respect to the {\it physical} momenta $\bold k$ in order to capture the response from conjugate physical fields.

The kinks from $\kappa_x(\bold k)$ and $\kappa_y(\bold k)$ correspond to discontinuities in their gradients, and hence also the Berry curvature. Take, for instance, the contribution
\begin{equation}
\frac{d D({\bar{k}_{x},\bar{k}_{y}})}{d k_{x}}=
i\left[\frac{\partial D({\bar{\bold k}})}{\partial \bar{k}_{x}}\frac{d \kappa_{x}(\bold k)}{d k_{x}}\!+\!\frac{\partial D({\bar{\bold k}})}{\partial \bar{k}_{y}}\frac{d \kappa_{y}(\bold k)}{d k_{x}}\right]\!.\label{nhbr}
\end{equation}
Even though $\partial D(\bar{\bold k})/\partial \bar{k}_{\mu}$, $\mu=x,y$ are in general continuous for gapped bands, they are multiplied by $d\kappa_\mu(\bold k)/dk_{x}$ gradients, which are discontinuous at the kinks.

For concreteness, we specialize Eq.~\eqref{eq:H} to a model that can be feasibly realized in an ultracold atomic lattice, as discussed later:
\begin{equation}\label{eq:d}
\begin{aligned}
&d_{x}=\Omega_{r} + 2\Omega_{0}\sin\phi_{0}\sin k_{x},\\
&d_{y}=2\Omega_{0}\cos\phi_{0}\sin k_{x},\\
&d_{z}=i\gamma - 2t_{x}\cos k_{x} - 2t_{y}\cos k_{y},
\end{aligned}
\end{equation}
where $t_{x}=\Omega_{0}$, $i\gamma$ originates from onsite loss under constant energy shift~\cite{li2019observation,lapp2019engineering,ren2022chiral}, and reciprocal couplings are between like pseudospins of neighboring unit cells. The detailed derivation for Eq.~\eqref{eq:d} can be found in Appendix \ref{Supp_1}.

To derive its kinked spectral deformation, we first diagonalize ${\cal H}(k_{x},k_{y}) $ via $\text{Det}({\cal H}-E\,\mathbb{I})=0$ to obtain the dispersion relation
\begin{equation}
E^{2}(k_{x},k_{y})=t_{+}z + t_{-}z^{-1} + t_{0},\label{eq:E2}
\end{equation}
where $z=e^{ik_x}$, $t_{+}= -2it_{x}(\gamma + \Omega_{r}\sin\phi_{0} + 2it_{y}\cos k_{y})$, $t_{-}= - 2it_{x}(\gamma -\Omega_{r}\sin\phi_{0} + 2it_{y}\cos k_{y})$, $t_{0}=\Omega_{r}^{2} - \gamma^{2} + 4t_{x}^{2} + 4t_{y}^{2}\cos^{2} k_{y} - 4i\gamma t_{y}\cos k_{y}$. Up to a constant $t_0$, Eq.~\eqref{eq:E2} reduces to Eq.~\eqref{eq:H1D} with $\alpha=\beta=1$, such that $\kappa_{x}$ is independent of $k_x$. As such, under $x$-OBCs and $y$-PBCs (cylinder), the effective Hamiltonian is ${\cal H}(\bar k_{x},k_{y})={\cal H}(k_{x}+i\kappa_{x}({\bf k}),k_{y})={\cal H}(k_{x}+i\kappa_x(k_{y}),k_{y})$, where
\begin{equation}
\kappa_x(k_{y}) \!=\! \ln\sqrt{\frac{t_{+}}{t_{-}}} \!=\! \ln\sqrt{\frac{\gamma \!+\! \Omega_{r}\sin\phi_{0} \!+\! 2it_{y}\cos k_{y}}{\gamma \!-\! \Omega_{r}\sin\phi_{0} \!+\! 2it_{y}\cos k_{y}}}.\label{eq:kx}
\end{equation}
Band singularities occur at non-analytic $k_{y}=\pm \frac{\pi}{2}$ under $\gamma<\Omega_{r}$ where $\text{Im}[\kappa_x(k_{y})]$ for ${\cal H}(\bar k_{x},k_{y})$ vanishes. Indeed, the real band surfaces $\text{Re}(E_\text{OBC})$ fracture along these lines [Fig.~\ref{fig:E_model_phy}(b)], and the real Berry curvature $\text{Re}[\bar\Omega_{xy}(\bold k)]$ likewise exhibits discontinuities [Fig.~\ref{fig:E_model_phy}(c)]. Due to the emergent non-locality, these discontinuities are expected to appear in other quantities characterizing the OBC band geometry or topology~\cite{marzari1997maximally}, such as the Fubini-Study metric~\cite{neupert2013measuring,lee2014lattice,lee2017band,mera2021engineering}.

\begin{figure}
\centering
\includegraphics[width=\linewidth]{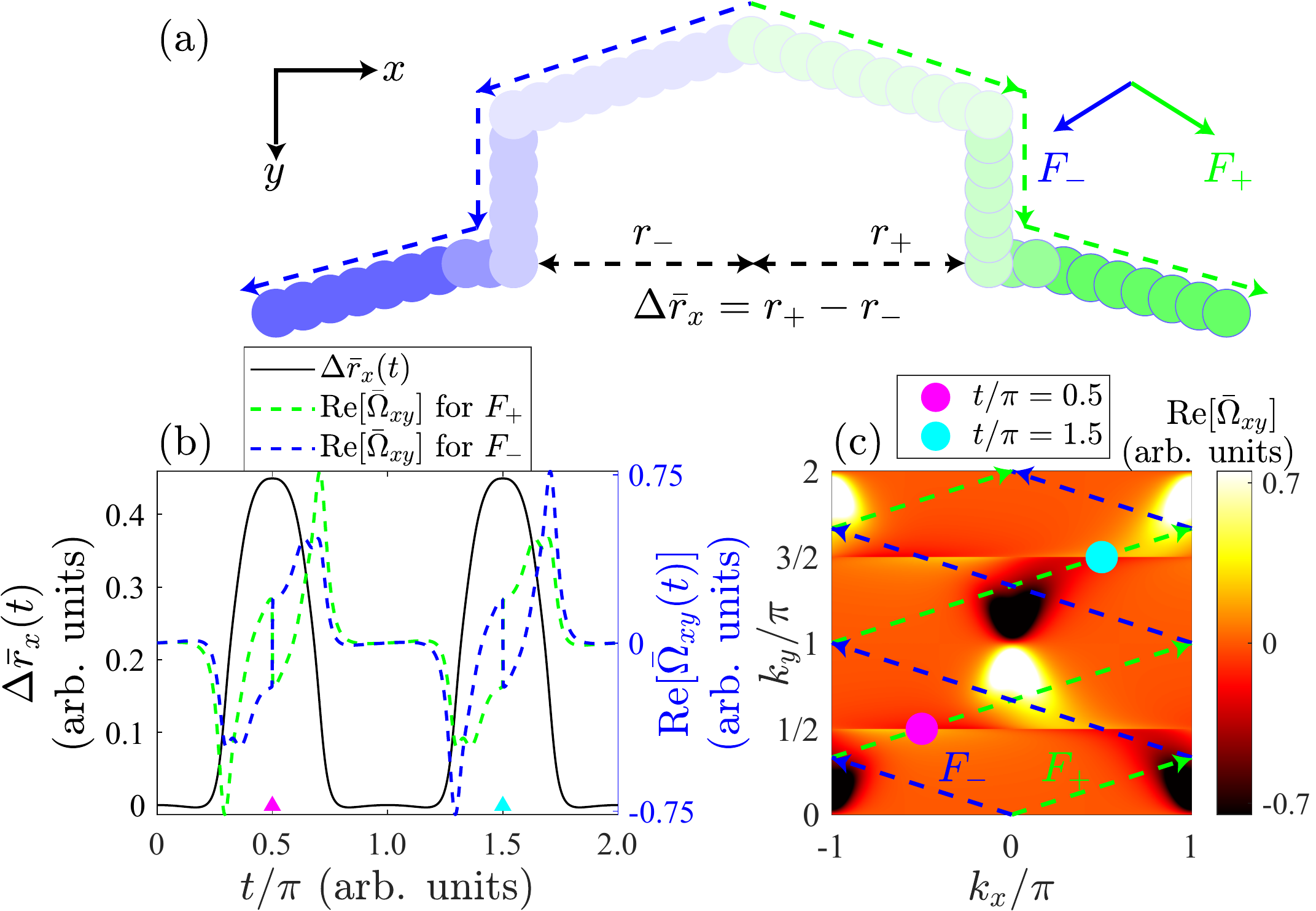}
\caption{\textbf{Kinked semiclassical responses from Berry curvature discontinuities.} (a) Schematics of detecting kinked responses through semiclassical dynamics. Two atomic clouds (green, blue) with zero initial momenta are respectively subject to forces ${\bf F}_{\pm}=F_{0}{\bf e}_{y}\pm F_{x}{\bf e}_{x}$, and consequently move along trajectories (dashed) related by reflection symmetry about $x=0$. The Berry curvature response can be extracted from their difference in $x$-center-of-mass $\Delta\overline{\mathbf{r}}_{x}$ [Eq.~\eqref{eq:rdif}]. (b) Due to non-Hermitian pumping, kinks appear in the $\Delta\overline{\mathbf{r}}_{x}$ (solid black) of our cold-atom model Eq.~\eqref{eq:d}. They coincide with the Berry curvature discontinuities (dashed) encountered during the evolution, which are driven by the force parameters $F_0=1$, $F_{x}=3$. (c) Origin of the response kinks, as seen from the two atomic cloud trajectories (green, blue) ${{\mathbf{k}}}_{\pm}(t)$. Evidently, Berry curvature discontinuities (beige/brown interfaces) are encountered at $F_{0}t/(\hbar/a_{y})=\pi/2,3\pi/2,...$ (blue and green dots), and they cause kinks to appear in $\Delta\overline{\mathbf{r}}_{x}$, as shown in (b). Parameters are $\hbar=1$, $a_{y}=1$, $\phi_{0}=\pi/4$, $t_{x}=1$, $t_{y}=t_{x}$, $\Omega_{r}=\Omega_{0}=t_{x}$, and $\gamma=0.5t_{x}$. Without loss of generality, these parameters are in arbitrary units (arb. units).}\label{fig:Trajectory}
\end{figure}

\section{Response kinks from discontinuous Berry curvature}\label{4}

We next demonstrate that the Berry curvature discontinuities [Fig.~\ref{fig:E_model_phy}(c)] are not just mathematical artifacts but are in fact experimentally detectable via semiclassical wave packet dynamics. In Hermitian systems, a wave packet can always be decomposed into real momentum eigenstates obeying well-known semiclassical equations of motion.
Such a picture remains valid in the presence of non-Hermitian pumping if the canonical momentum is replaced by the complex $\kappa$-deformed momentum.
The velocity of a wave packet in a generic bounded 2D non-Hermitian system evolves as~\cite{longhi2009bloch,longhi2015bloch,xu2017weyl}
\begin{equation}
\overline{\mathbf{v}}(t)\!=\!\frac{d\bar{\bf r}}{dt} \!=\! \frac{1}{\hbar}\frac{\partial {\rm Re}[E(\bar{\bf k})]}{\partial{\bf k}} \!-\! \frac{1}{\hbar}({\bf F}\times{\bf e}_{z}){\rm Re}[\bar{\Omega}_{xy}(\bar{\bf k})],\label{eq:dynamics0}
\end{equation}
where $\bar{\bf r}$ describes its center of mass, and ${\bf F}(t) = (F_x(t), F_y(t))$ is an externally applied force.
The first term $\partial {\rm Re}[E(\bar{\bf k})]/\partial{\bf k}$ is the usual canonical velocity, with $E(\bar{\bf k})$ being that of the lower occupied band, but evaluated at the effective momentum $\bar{\bf k}$. The second term involving ${\rm Re}[\bar{\Omega}_{xy}(\bar{\bf k})]$ is the anomalous velocity contribution.

To isolate the Berry curvature response and extract signatures of its discontinuities, we compare the semiclassical trajectories of two initially stationary wave packets subject to forces ${\bf F}_{\pm}=F_{y}(t){\bf e}_{y}\pm F_{x}{\bf e}_{x}$ with opposite ${\bf e}_x$ components, as sketched in Fig.~\ref{fig:Trajectory}(a). These wave packets acquire different effective momenta~\cite{price2012mapping,wimmer2017experimental} $\bar{\bf k}_{\pm}(t)={\bf k}_{\pm}(t) + i\kappa_{x}({\bf k}_{\pm}(t))$, where ${\mathbf{k}}_{\pm}(t)=\hbar^{-1}\int_{0}^{t}{\bf F}_{\pm}dt^{\prime}$, and thus propagate along two different trajectories. If we further assume an even energy dispersion, such that $\frac{\partial {\rm Re}[E(\overline{\mathbf{k}}_{+}(t))]}{\partial k_{x}}+\frac{\partial{\rm Re}[E(\overline{\mathbf{k}}_{-}(t))]}{\partial k_{x}}=0$ (Appendix \ref{Supp_2}), only the Berry curvature contribution survives, and the two trajectories are separated by an $x$ displacement of
\begin{equation}
\begin{aligned}
\Delta\overline{\mathbf{r}}_{x}(t)\!=\!&\int_{0}^{t}\left[\overline{\mathbf{v}}_{x}(t^{\prime})|_{{\bf F}_{+}}\!+\!\overline{\mathbf{v}}_{x}(t^{\prime})|_{{\bf F}_{-}}\right]\!dt^{\prime} \\
\!=\!&\!-\!\frac{1}{\hbar}\!\int_{0}^{t}\!\!\!F_{y}(t\rq{})\!\operatorname{Re}\!\left[\bar{\Omega}_{xy}\left(\overline{\mathbf{k}}_{-}\!\left(t^{\prime}\right)\right)\!+\!\bar{\Omega}_{xy}\left(\overline{\mathbf{k}}_{+}\!\left(t^{\prime}\right)\right)\!\right]\!d t^{\prime}.\label{eq:rdif}
\end{aligned}
\end{equation}

Plotted in Fig.~\ref{fig:Trajectory}(b) is the evolution of the center-of-mass difference $\Delta \overline{\mathbf{r}}_{x}(t)$ (solid black) for our model Eq.~\eqref{eq:d}, it exhibits sharp kinks when discontinuities are encountered in the underlying Berry curvatures ${\rm Re}[\bar{\Omega}_{xy}(\bar{\bf k}_{\pm}(t))]$ (green, blue) along the trajectories. This is concretely illustrated in Fig.~\ref{fig:Trajectory}(c), which shows the momentum-space trajectories ${\bf k}_{\pm}(t)$ (green, blue) on the ${\rm Re}[\bar{\Omega}_{xy}({\bf k}_{\pm})]$ landscape from Fig.~\ref{fig:E_model_phy}(c). With constants $F_{x}$ and $F_{y}(t)=F_{0}$ such that ${\bf k}_{\pm}(t)$ traces out the paths $k_{x}=\pm(F_{x}/F_{0})k_{y}$, Berry curvature discontinuities are encountered by both trajectories whenever $\hbar k_{y}=F_{0}t=(\hbar/a_{y})(n+1/2)\pi$. These discontinuities exert sudden impulses that give rise to abrupt kinks in the real-space trajectory, as captured by $\Delta \overline{\mathbf{r}}_{x}$. These enigmatic kinks, which arise solely from the non-locality of non-Hermitian pumping and not physical impulses, can be detected through the motion of  atoms in our designed experimental setup below.

\begin{figure}
\centering
\includegraphics[width=\linewidth]{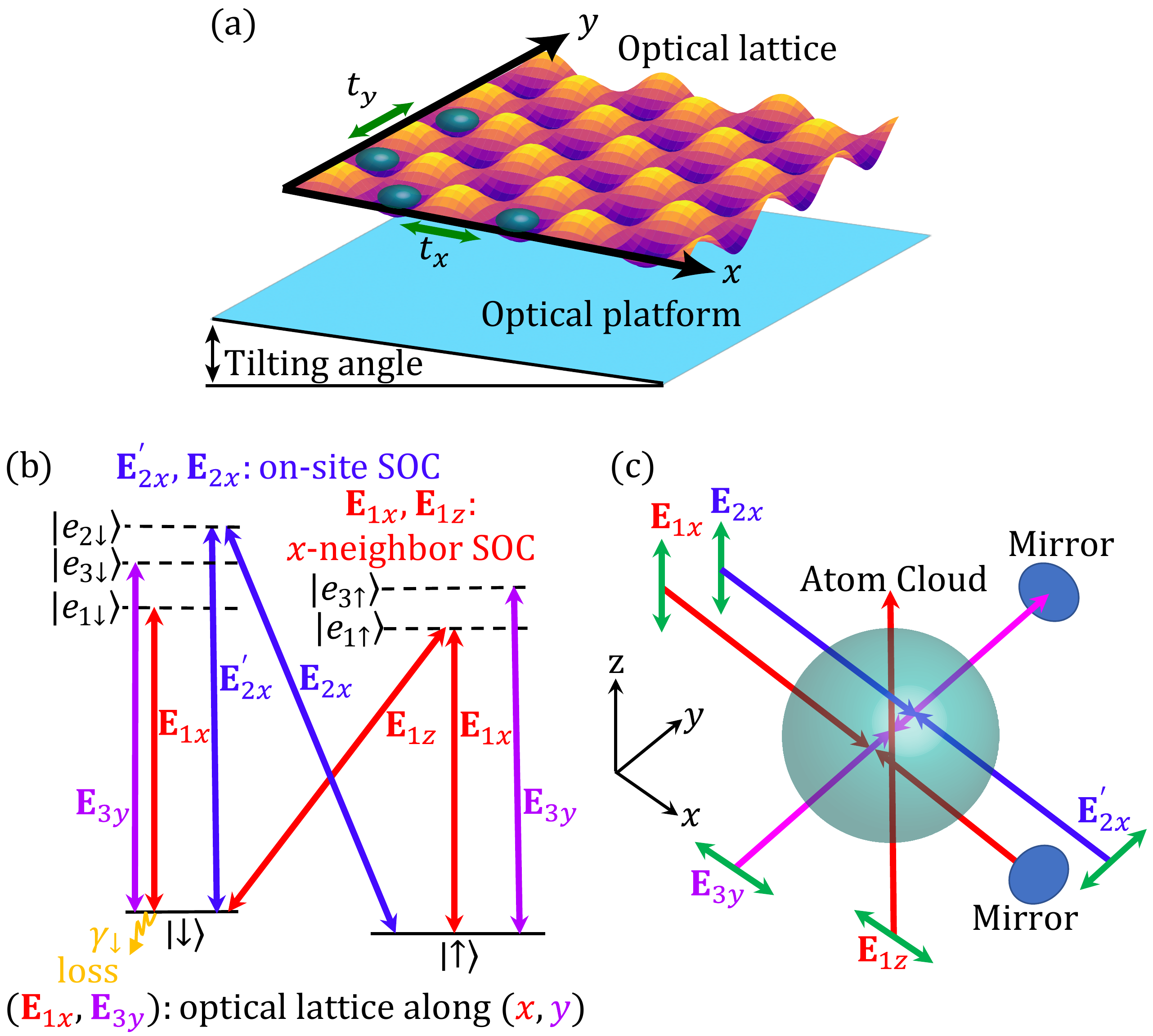}
\caption{\textbf{Ultracold atomic setup.} (a) Cold atoms (cyan) are trapped in a tilted optical lattice with effective hoppings $t_{x}$, $t_{y}$ created by the lasers described in panels (b) and (c). They move according to Eq.~\eqref{eq:dynamics0}, subject to a force ${\bf F}$ with $x$ and $y$ components, respectively, arising from tilting and shaking the optical lattice. (b) To realize our Hamiltonian in Eq.~\eqref{eq:d}, various Raman lasers generate effective onsite (blue) and nearest-neighbor (red) spin-orbit couplings (SOCs) between $|\uparrow\rangle$ and $|\downarrow\rangle$ ground states, mediated by excited states $|e_{2\downarrow}\rangle$ and $|e_{1\uparrow}\rangle$. Other lasers partake in generating the optical lattice potential (red, purple) and laser-induced loss (yellow). (c) Three-dimensional (3D) configuration of the lasers in (b), with polarization directions indicated in green.}\label{fig:Schematics}
\end{figure}

\section{Detecting kinked responses with cold atoms}\label{5}

Ultracold atomic platform simulations for novel condensed-matter phenomena have gained significant attention recently~\cite{li2019observation,lapp2019engineering,ren2022chiral,liang2022dynamic,gou2020tunable}. Our proposed cold-atom setup showcases the potential of synthetic spin-orbit coupling~\cite{lin2011spin,huang2016experimental,wu2016realization,wang2021realization}, laser-induced loss~\cite{ren2022chiral}, and optical lattice potential~\cite{atala2013direct,goldman2016topological,yang2020cooling} as powerful tools for quantum control and simulations.

We propose a cold-atom setup [Figs.~\ref{fig:Schematics}(a)-\ref{fig:Schematics}(c)] for realizing the model Eq.~\eqref{eq:d} with predicted response kinks. It requires atoms with two ground states (labeled as $|\uparrow\rangle,|\downarrow\rangle$) and five suitably located excited states [Fig.~\ref{fig:Schematics}(b)], such as $^{87}$Rb, whose detailed energy levels are elaborated in Appendix \ref{Supp_1}. Several Raman lasers connect the two $|\uparrow\rangle,|\downarrow\rangle$ ground states through various excited states $|e_{1\uparrow}\rangle,|e_{2\downarrow}\rangle$. Laser fields ${\bf E}_{1x}$ (red) and ${\bf E}_{3y}$ (purple), respectively, generate the optical lattice potential in the ${\bf e}_x$ and ${\bf e}_y$ directions, which also determine the nearest-neighbor hopping amplitudes $t_{x}$, $t_{y}$ in the effective tight-binding description (Appendix \ref{Supp_1}). Pairs of Raman laser fields ${\bf E}_{2x}, {\bf E}'_{2x}$ (blue) and ${\bf E}_{1x},{\bf E}_{1z}$ (red) respectively induce effective on-site and $x$-nearest-neighbor spin-orbit coupling terms $\Omega_{r}$ and $\Omega_{0}$, mediated by excited states $|e_{2\downarrow}\rangle$ and $|e_{1\uparrow}\rangle$. A reciprocity-breaking phase $\phi_{0}$ can be introduced through Ramam laser beams ${\bf E}_{1x}$ and ${\bf E}_{1z}$. Finally, non-Hermiticity is introduced through laser-induced loss with rate $\gamma_{\downarrow}$ by depopulating atoms from the ground state $|\downarrow\rangle$ to another excited state~\cite{li2019observation,lapp2019engineering,ren2022chiral,zhou2022engineering,li2022dissipative,shen2023proposal}. All in all, as detailed in Appendix \ref{Supp_1}, we arrive at the effective tight-binding Hamiltonian given by Eq.~\eqref{eq:d} in momentum space, with the additional restriction $\Omega_{0}=t_{x}$ so that the Berry curvature is discontinuous. Related experimental setups have demonstrated considerable freedom in tuning $t_{x}/\hbar$, $t_{y}/\hbar$, $\Omega_{0}/\hbar$, and $\Omega_{r}/\hbar$ within the range of $2\pi\times[60,1000]$ Hz~\cite{li2019observation,lapp2019engineering,liang2022dynamic,ren2022chiral,zhou2022engineering,folling2007direct,lin2011spin,atala2013direct,huang2016experimental,wu2016realization}. The required $x$ direction OBCs can be realized by lifting the optical potential along the edges~\cite{zhou2022engineering}.

\begin{figure}
\centering
\includegraphics[width=\linewidth]{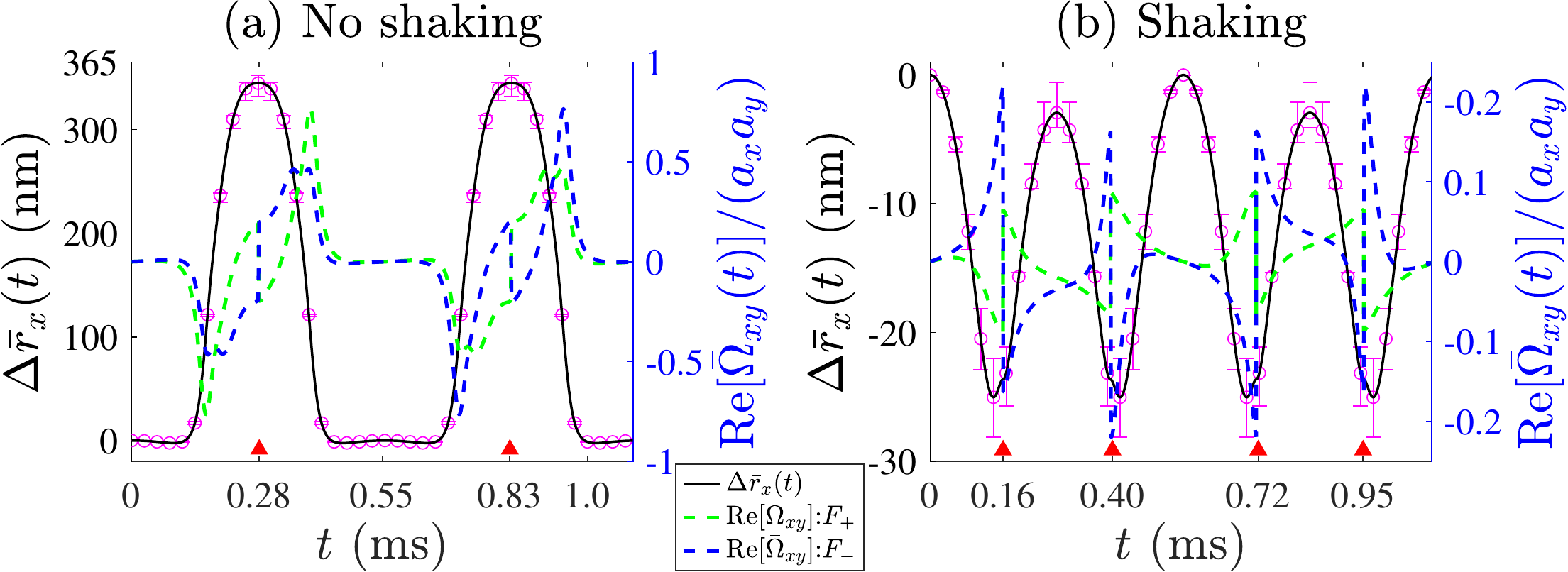}
\caption{\textbf{Simulated $x$-displacement differences $\Delta\overline{\mathbf{r}}_{x}(t)$ between two atomic clouds under opposite $F_x$ tilts, which generically exhibit prominent response kinks without fine-tuning of parameters.} (a) No shaking: $F_{x}a_{x}/t_{x}=3$ and $F_{0}a_{y}/t_{x}=1$. (b) Shaking: $F_{x}a_{x}/t_{x}=F_{0}a_{y}/t_{x}=2$ and the shaking frequency $\omega=t_{x}/\hbar$. The data with error bars are calculated with the fluctuations of the relative phase $\phi_{0}$, which takes a uniform distribution over $[\pi/4-\delta\phi,\pi/4+\delta\phi]$ with $\delta\phi=0.03\pi$. Other parameters are $a_{x}=a_{y}=767$ nm, $t_{x}/\hbar=2\pi\times900$ Hz, $t_{y}=t_{x}$, $\Omega_{r}=\Omega_{0}=t_{x}$, and $\gamma=0.5t_{x}$.}\label{fig:Simulations}
\end{figure}

To observe the response kinks, one can initialize two independent groups of atoms with zero momenta~\cite{anderson1995observation} and drive the two atom clouds~\cite{aidelsburger2015measuring,mancini2015observation,goldman2016topological,cooper2019topological} with different periodic force fields ${\bf F}_{\pm}(t)=F_{0}\cos(\omega t){\bf e}_{y}\pm F_{x}{\bf e}_{x}$. We choose an oscillatory $F_{y}(t)=F_{0}\cos(\omega t)$, which can be implemented through shaking~\cite{parker2013direct,jotzu2014experimental}, to confine the atomic clouds to a small sample region. The $F_{x}$ force component can be realized by tilting the optical platform.

Through density snapshots~\cite{aidelsburger2015measuring,goldman2016topological}, response kinks can be detected in the difference in the $x$ center of mass $\Delta\overline{\mathbf{r}}_{x}(t)$ of the two clouds, as plotted in Fig.~\ref{fig:Simulations}(a) for an experimentally realistic scenario characterized by $t_{x}/\hbar=2\pi\times 900$ Hz, forces $F_{x}=3t_{x}/a_{x}$ and $F_{0}=t_{x}/a_{y}$, and lattice constants $a_x=a_y=767$ nm~\cite{su2023observation,yang2020cooling,yang2017spin}. $F_x$ can be gravitationally introduced through a tilt angle [Fig.~\ref{fig:Schematics}(a)] of $\arcsin\left[t_{x}/(mga_{x})\right]$, which is $\approx33^{\circ}$ for $^{87}$Rb atoms of mass $m\approx1.45667\times10^{-25}$ kg.
Importantly, since the response kinks arise inevitably whenever the semi-classical trajectories cross Berry curvature discontinuities, they are observable without fine-tuning, as evident in the approximately responses from the slight fluctuations of the relative phase $\phi_{0}$, which takes a uniform distribution over [$\pi/4-\delta\phi,\pi/4+\delta\phi$] with $\delta\phi=0.03\pi$. In Fig.~\ref{fig:Simulations}(a), we use constant force without shaking. The kinks (abrupt jumps) in $\Delta\overline{\mathbf{r}}_{x}(t)$ occur at times $t=0.28$ and $0.83$ ms (red triangles) whenever $k_{y}(t)a_{y}$ crosses $\pi/2$ and $3\pi/2$, where Berry curvature discontinuities occur. In Fig.~\ref{fig:Simulations}(b), for comparison, we use periodic shaking force with shaking frequency. The kinks in $\Delta\overline{\mathbf{r}}_{x}(t)$ occur at times $t=0.16$, $0.40$, $0.72$, and $0.95$ ms (red triangles) whenever $k_{y}(t)a_{y}$ crosses $\pm\pi/2$, where Berry curvature discontinuities occur.
With response kinks occurring within the timescale of 1 ms, decoherence will not be significant~\cite{li2019observation,liang2022dynamic}. Besides, near the kinks, the error bars become larger in both Figs.~\ref{fig:Simulations}(a) and \ref{fig:Simulations}(b).

\section{Discussion}\label{6}

In this work, we uncovered intriguing non-Hermitian kinked responses beyond the non-Hermitian skin effect, as captured by discontinuities in the Berry curvature, or in fact, any band geometric quantity containing momentum derivatives of the band eigenstates. The advantage of our experimental proposal is the capacity to measure the signature of kinked responses as prominent kinks in short-time dynamics, which is tailored to the state-of-the-art cold atom experimental platforms. Our proposal lays the groundwork for future experimental demonstrations of other non-Hermitian transport phenomena with ultracold atomic platforms.

Our kinks are distinct from collective-excitation-induced kinks observed in harmonically trapped quantum gases~\cite{jin1996collective,mewes1996collective,chevy2002transverse,kinoshita2006quantum,lobser2015observation}, or finite-lattice-mismatch-induced kinks associated with the commensurate-incommensurate transition in cold atoms~\cite{buchler2003commensurate,gangloff2020kinks}.

Curiously, even though the discontinuous Berry curvature is phenomenologically manifested in the semi-classical response of wave packets, it does not affect the Hall response of the {\it entire} band. Indeed, the Chern number $
\bar{{\cal C}}=-\frac{1}{2\pi}\int{\rm Re}(\bar{\Omega}_{xy})dk_{x}dk_{y}$ can be shown to surprisingly remain quantized despite the violent deformations of the Brillouin zone.

All data and code for this work are available from the corresponding authors upon reasonable request.

\begin{acknowledgments}
We acknowledge helpful discussions with Haowei Li and Lihong Zhou. C.H.L. and F.Q. acknowledge support from the QEP2.0 Grant from the Singapore National Research Foundation (Grant No. NRF2021-QEP2-02-P09) and the Singapore Ministry of Education Academic Research Fund  Tier-II Grant (Award No. MOE-T2EP50222-0003).
\end{acknowledgments}


\appendix
\onecolumngrid







\section{Derivations for the analytical expression for $E_{\rm OBC}(k_{x})$ below Eq.~\eqref{eq:H1Dkappa}}\label{Supp_0}

In this Appendix, we will show the detailed derivations for the analytical expression for $E_{\rm OBC}(k_{x})$ below Eq.~\eqref{eq:H1Dkappa}.

With Eq.~\eqref{eq:H1Dkappa}, the solution for the GBZ, i.e., $z(k_{x})$, is given by
\begin{eqnarray}
z(k_x) =&
e^{i\frac{2\pi m}{\alpha+\beta}}e^{-\kappa(k_x)}e^{ik_{x}} 
=e^{i\left(k_{x}+\frac{2\pi m}{\alpha+\beta}\right)}\left(\frac{A\sin(\alpha k_x)}{B\sin(\beta k_x)}\right)^{-\frac{1}{\alpha+\beta}} 
=e^{i\left(k_{x}+\frac{2\pi m}{\alpha+\beta}\right)}\left(\frac{B\sin(\beta k_x)}{A\sin(\alpha k_x)}\right)^{\frac{1}{\alpha+\beta}}.
\end{eqnarray}
Furthermore, the analytical expression for $E_{\rm OBC}(k_{x})$ below Eq.~\eqref{eq:H1Dkappa} can be derived as
\begin{eqnarray}
E_\text{OBC}(k_x) &\!=\!& Az^\alpha+Bz^{-\beta} \nonumber\\
&\!=\!&
Ae^{i\alpha\left(k_{x}+\frac{2\pi m}{\alpha+\beta}\right)}\left[\frac{B\sin(\beta k_x)}{A\sin(\alpha k_x)}\right]^{\frac{\alpha}{\alpha+\beta}} 
+Be^{-i\beta\left(k_{x}+\frac{2\pi m}{\alpha+\beta}\right)}\left[\frac{A\sin(\alpha k_x)}{B\sin(\beta k_x)}\right]^{\frac{\beta}{\alpha+\beta}} \nonumber\\
&\!=\!&
Ae^{i\alpha\left(k_{x}+\frac{2\pi m}{\alpha+\beta}\right)}\left[\frac{B^{\alpha}\sin^{\alpha}(\beta k_x)}{A^{\alpha}\sin^{\alpha}(\alpha k_x)}\right]^{\frac{1}{\alpha+\beta}} 
+Be^{-i\beta\left(k_{x}+\frac{2\pi m}{\alpha+\beta}\right)}\left[\frac{A^{\beta}\sin^{\beta}(\alpha k_x)}{B^{\beta}\sin^{\beta}(\beta k_x)}\right]^{\frac{1}{\alpha+\beta}} \nonumber\\
&\!=\!&
Ae^{i\alpha\left(k_{x}+\frac{2\pi m}{\alpha+\beta}\right)}\left[\frac{\sin(\beta k_x)}{A}\right]^{\frac{\alpha}{\alpha+\beta}}\left[\frac{B^{\alpha}}{\sin^{\alpha}(\alpha k_x)}\right]^{\frac{1}{\alpha+\beta}} 
+Be^{-i\beta\left(k_{x}+\frac{2\pi m}{\alpha+\beta}\right)}\left[\frac{\sin(\alpha k_x)}{B}\right]^{\frac{\beta}{\alpha+\beta}}\left[\frac{A^{\beta}}{\sin^{\beta}(\beta k_x)}\right]^{\frac{1}{\alpha+\beta}} \nonumber\\
&\!=\!&\left[\frac{A^{\beta}}{\sin^{\beta}(\beta k_x)}\right]^{\frac{1}{\alpha+\beta}}\left[\frac{B^{\alpha}}{\sin^{\alpha}(\alpha k_x)}\right]^{\frac{1}{\alpha+\beta}}\left\{
Ae^{i\alpha\left(k_{x}\!+\!\frac{2\pi m}{\alpha+\beta}\right)}\left[\frac{\sin(\beta k_x)}{A}\right]^{\frac{\alpha+\beta}{\alpha+\beta}} 
\!+\!Be^{-i\beta\left(k_{x}\!+\!\frac{2\pi m}{\alpha+\beta}\right)}\left[\frac{\sin(\alpha k_x)}{B}\right]^{\frac{\alpha+\beta}{\alpha+\beta}} \right\} \nonumber\\
&\!=\!&\left[\frac{A^{\beta}B^{\alpha}}{\sin^{\alpha}(\alpha k_x)\sin^{\beta}(\beta k_x)}\right]^{\frac{1}{\alpha+\beta}} \left[
e^{i\alpha\left(k_{x}+\frac{2\pi m}{\alpha+\beta}\right)}\sin(\beta k_x)
+e^{-i\beta\left(k_{x}+\frac{2\pi m}{\alpha+\beta}\right)}\sin(\alpha k_x) \right] \nonumber\\
&\!=\!&\left[\frac{A^{\beta}B^{\alpha}}{\sin^{\alpha}(\alpha k_x)\sin^{\beta}(\beta k_x)}\right]^{\frac{1}{\alpha+\beta}} \frac{1}{2i}\left[
e^{i\alpha\left(k_{x}+\frac{2\pi m}{\alpha+\beta}\right)}(e^{i\beta k_x} \!-\! e^{-i\beta k_x})
\!+\!e^{-i\beta\left(k_{x}+\frac{2\pi m}{\alpha+\beta}\right)}(e^{i\alpha k_x} \!-\! e^{-i\alpha k_x}) \right] \nonumber\\
&\!=\!&\left[\frac{A^{\beta}B^{\alpha}}{\sin^{\alpha}(\alpha k_x)\sin^{\beta}(\beta k_x)}\right]^{\frac{1}{\alpha\!+\!\beta}} \frac{1}{2i}\left[
e^{i\left[(\alpha\!+\!\beta)k_{x}\!+\!\frac{2\pi m\alpha}{\alpha+\beta}\right]} \!-\! e^{i\left[(\alpha\!-\!\beta)k_{x}\!+\!\frac{2\pi m\alpha}{\alpha+\beta}\right]}
\!+\!e^{i\left[(\alpha\!-\!\beta)k_{x}\!-\!\frac{2\pi m\beta}{\alpha+\beta}\right]} \!-\! e^{-i\left[(\alpha\!+\!\beta)k_{x}\!+\!\frac{2\pi m\beta}{\alpha+\beta}\right]} \right] \nonumber\\
&\!=\!&\left[\frac{A^{\beta}B^{\alpha}}{\sin^{\alpha}(\alpha k_x)\sin^{\beta}(\beta k_x)}\right]^{\frac{1}{\alpha+\beta}} e^{i\frac{2\pi m\alpha}{\alpha+\beta}}\frac{1}{2i}\left[
e^{i(\alpha+\beta)k_{x}} - e^{i(\alpha-\beta)k_{x}}
+e^{i(\alpha-\beta)k_{x}} - e^{-i(\alpha+\beta)k_{x}} \right] \nonumber\\
&\!=\!&\left[\frac{A^{\beta}B^{\alpha}}{\sin^{\alpha}(\alpha k_x)\sin^{\beta}(\beta k_x)}\right]^{\frac{1}{\alpha+\beta}} e^{i\frac{2\pi m\alpha}{\alpha+\beta}}\sin\left[(\alpha+\beta)k_{x}\right].
\end{eqnarray}

\section{Effective non-Hermitian model for discontinuous Berry curvature and its realization in an ultracold atomic setup}\label{Supp_1}

In the main text, to explore the kinked response from the non-Hermitian pumping~\cite{yao2018edge,okuma2020topological,li2020critical}, we require a lattice model that exhibits discontinuous Berry curvature. Here, we show how such an effective lattice model can result from a cold-atom optical lattice setup described by the following Hamiltonian density
\begin{equation}
\mathcal{H}(x,y)\!=\!\frac{\hat{\bf p}^2}{2m} \!+\!V(x,y) \!+\! [({\rm A}(x)\!+\!{\rm B}(x))\sigma_{+} \!+\! {\rm h.c.}] \!+\! i\gamma\sigma_{z},\label{eq:suppHeff}
\end{equation} in a two-component basis $(\left| \uparrow\right\rangle\!,\!\left| \downarrow\right\rangle)^{T}$, taken from appropriate atomic states described below. Here $\hat{\bf p}^2=\hat{p}_{x}^2+\hat{p}_{y}^2$ is the squared momentum operator, ${\rm A}(x)$ and ${\rm B}(x)$ are the spin-orbit couplings (SOCs) induced by two sets of Raman lasers that generate the on-site and  nearest-neighbor spin flip, $\gamma$ is the effective decay rate of the spin-dependent on-site loss achieved by laser, and $\sigma_{\pm}=(\sigma_{x}\pm i\sigma_{y})/2$ with the Pauli matrices $\sigma_{j}$ ($j=x,y,z$). The optical lattice potential is of the form
\begin{equation}
V(x,y)=- V_{x}\cos^{2}(\pi x/a_{x}+\phi_{1-}) - V_{y}\cos^{2}(\pi y/a_{y}+\phi_{3-}),
\end{equation}
where $V_x$ and $V_y$ are the lattice depths, $a_{x}$ and $a_{y}$ are lattice constants~\cite{wu2016realization}, and phases $\phi_{1-}$ and $\phi_{3-}$ are determined by the electric fields ${\bf E}_{1x}$ and ${\bf E}_{3y}$ that will be described later in section \ref{Supp_1.4}.

Below, we elaborate on the details of each aspect of the above experimental setup.

\subsection{Raman-transition scheme for $^{87}$Rb}\label{Supp_1.1}

\begin{figure}[h]
\centering
\includegraphics[width=0.4\textwidth]{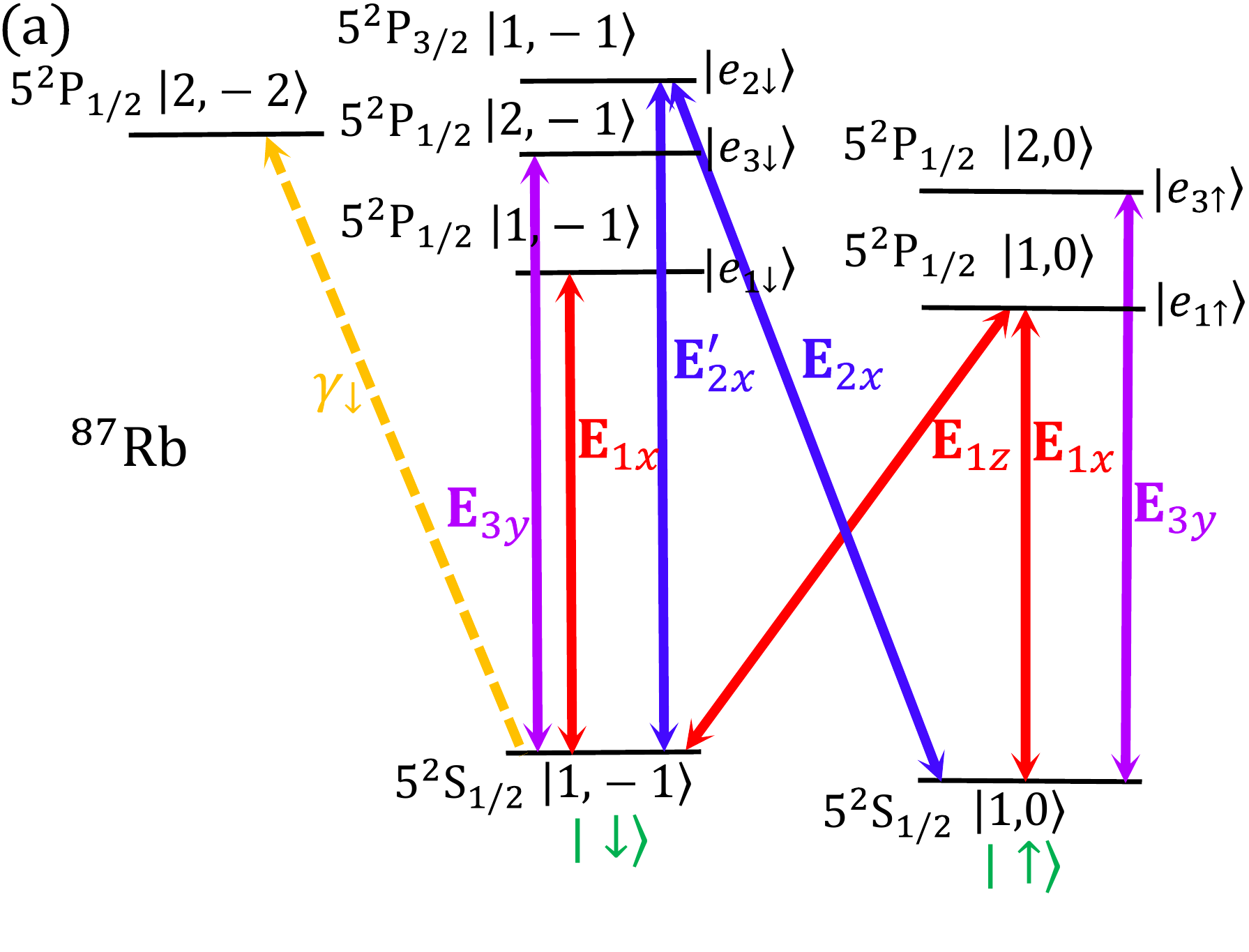}
\includegraphics[width=0.66\textwidth]{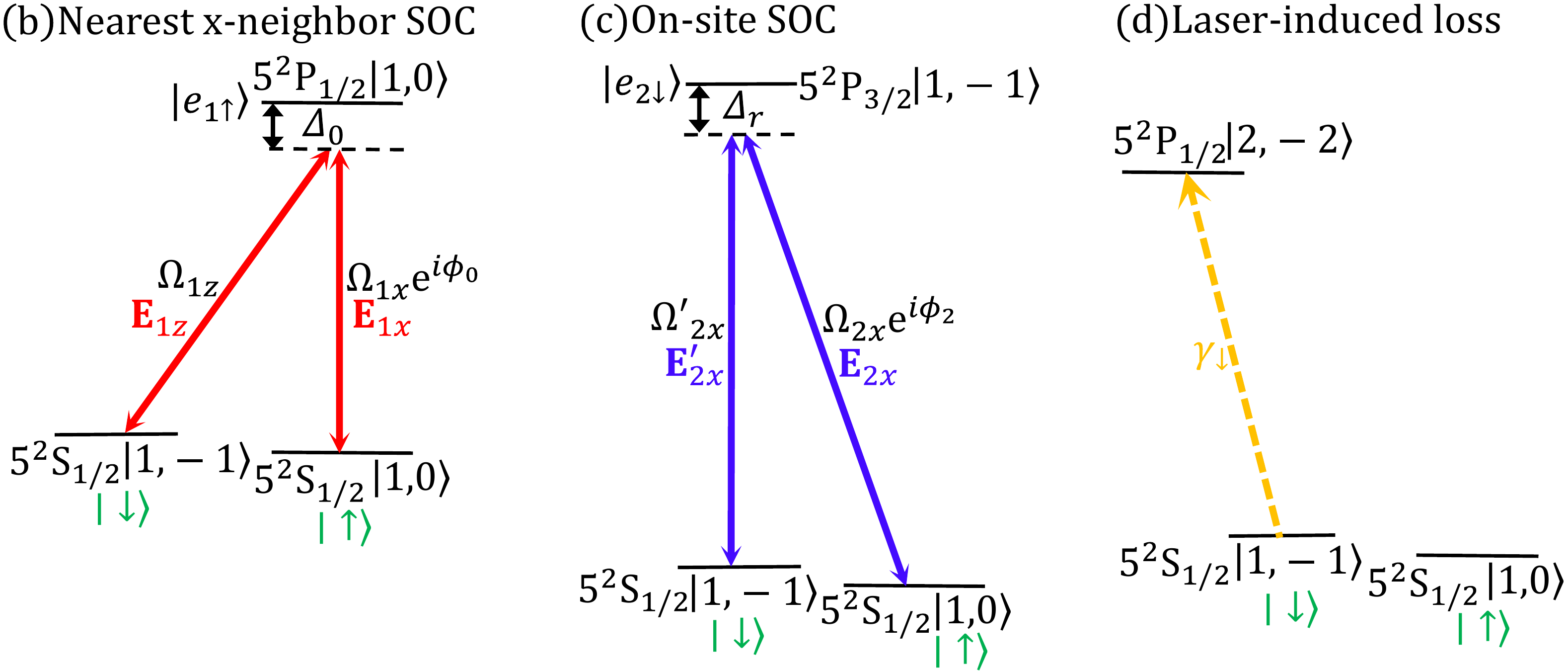}
\caption{(a) Energy-level structures for the relevant Raman transitions in $^{87}$Rb. Red (${\bf E}_{1x},{\bf E}_{1z}$) and Blue (${\bf E}_{2x},{\bf E}\rq{}_{2x}$) arrows denote laser transitions for effective spin-orbit coupling (SOC) between the ground states, mediated by excited states. Red (${\bf E}_{1x}$) and purple (${\bf E}_{3y}$) arrows generate the $x$- and $y$-direction optical lattice potentials. Another laser (yellow) gives rise to depopulation losses from the $|\downarrow\rangle$ ground state. The chosen hyperfine ground states are $|\downarrow\rangle=|F=1,m_{F}=-1\rangle$ and $|\uparrow\rangle=|F=1,m_{F}=0\rangle$ for 5$^2$S$_{1/2}$, and the chosen excited states are $|e_{1\uparrow}^{}\rangle=|F\rq{}=1,m'_{F}=0\rangle$, $|e_{1\downarrow}^{}\rangle=|F\rq{}=1,m'_{F}=-1\rangle$, $|e_{3\uparrow}^{}\rangle=|F\rq{}=2,m'_{F}=0\rangle$, $|e_{3\downarrow}^{}\rangle=|F\rq{}=2,m'_{F}=-1\rangle$ for 5$^2$P$_{1/2}$, and $|e_{2\downarrow}^{}\rangle=|F\rq{}=1,m'_{F}=-1\rangle$ for 5$^2$P$_{3/2}$. (b) Raman transitions giving rise to the nearest $x$ neighbor SOC with amplitude $M_0\sin(k_1x)e^{i\phi_0}$ with $M_0=-\hbar\Omega_{1x}\Omega_{1z}/\Delta_0$, which is mediated by the $|e_{1\uparrow}\rangle=|F\rq{}=1,m'_{F}=0\rangle$ excited state. (c) Raman transitions that give rise to the on-site effective SOC with amplitude $M_r e^{i(2k_2x+\phi_2)}$ with $M_r=-\hbar\Omega_{2x}\Omega'_{2x}/\Delta_r$, which is mediated by the $|e_{2\downarrow}^{}\rangle=|F\rq{}=1,m'_{F}=-1\rangle$ excited state. (d) Laser-induced depopulation loss for the $i\gamma_{\downarrow}^{}\sigma_z$ term. We couple one of the ground states $|\downarrow\rangle$ with an excited state $|F\rq{}=2,m\rq{}_{F}=-2\rangle$ of 5$^2$P$_{1/2}$. The yellow dashed arrow line represents the laser-induced loss from Eq.~\eqref{eq:loss}.}\label{fig:Schematics_Rb87}
\end{figure}

To engineer the model Eq.~\eqref{eq:suppHeff} in a cold atom system, we propose to use the energy levels of $^{87}$Rb atoms as a concrete realization~\cite{wu2016realization,lin2011spin}. As shown in the Raman transition diagram in Fig.~\ref{fig:Schematics_Rb87}(a), there are two hyperfine ground states $\left|\uparrow\right\rangle=\left|F=1,m_{F}=0\right\rangle$ and $\left|\downarrow\right\rangle=\left|F=1,m_{F}=-1\right\rangle$ from the 5$^{2}$S$_{1/2}$ energy manifold. We can also make use of five additional hyperfine excited states: four in the 5$^{2}$P$_{1/2}$ manifold: $\left|F'=1,m'_{F}=0\right\rangle$, $\left|F'=1,m'_{F}=-1\right\rangle$, $\left|F'=2,m'_{F}=0\right\rangle$, $\left|F'=2,m'_{F}=-1\right\rangle$, and $\left|F'=1,m'_{F}=-1\right\rangle$ in the 5$^{2}$P$_{3/2}$ manifold.


\subsection{Two types of spin-orbit couplings}\label{Supp_1.2}

Here, we discuss the two types of spin-orbit couplings (SOCs) in our setup:\\

\begin{enumerate}
\item Type 1: Effective nearest $x$-neighbor SOC $M_0$ [shown again in Fig.~\ref{fig:Schematics_Rb87}(b)]
	
This is generated by two Raman lasers: a standing wave along the $x$ direction with the optical-field vector ${\bf E}_{1x}$ (which also generates the optical lattice $V_x$ along the $x$ direction), and a plane wave propagating along the $z$ direction with the optical-field vector ${\bf E}_{1z}$, as given by
\begin{eqnarray}
{\bf E}_{1x}&\!=\!&E_{1x}e^{i\phi_{1+}}[e^{i(k_{1}x+\phi_{1-})}\!+\!e^{-i(k_{1}x+\phi_{1-})}]{\bf e}_{z}\!=\!2E_{1x}e^{i\phi_{1+}}\cos(k_{1}x+\phi_{1-}){\bf e}_{z},\label{eq:E1x}\\
{\bf E}_{1z}&\!=\!&E_{1z}e^{ik_{1}\rq{}z+i\phi_{1z}}{\bf e}_{x},
\end{eqnarray} where ${\bf e}_{z}$ and ${\bf e}_{x}$ are the directions of the polarization, $\phi_{1\pm}=(\phi_{1x}\pm\phi'_{1x})/2$ with $\phi_{1x}$ and $\phi'_{1x}$ the phases of incident and reflected lights, respectively.
By fixing $\phi_{1-}=-\pi/2$ and choosing a frame such that $z=0$, we have
\begin{eqnarray}
{\bf E}_{1x}&\!=\!&2E_{1x}e^{i\phi_{1+}}\sin(k_{1}x){\bf e}_{z},\\
{\bf E}_{1z}&\!=\!&E_{1z}e^{i\phi_{1z}}{\bf e}_{x}.
\end{eqnarray}
The optical fields ${\bf E}_{1x}$ and ${\bf E}_{1z}$ give rise to transitions between the ground states $|\uparrow\rangle,|\downarrow\rangle$ and the excited state 5$^{2}$P$_{1/2}|1,0\rangle$ [Figs.~\ref{fig:Schematics_Rb87}(a) and \ref{fig:Schematics_Rb87}(b)], thereby leading to effective SOC between pseudospins $|\uparrow\rangle$ and $|\downarrow\rangle$ mediated by this excited state. By adiabatic elimination, it can be shown that~\cite{qin2023non} the SOC amplitude is given by $M_{0}\sin(k_{1}x)e^{i\phi_{0}}$, where $\phi_{0}=\phi_{1+}-\phi_{1z}$ and $M_{0}=-\hbar\Omega_{1x}\Omega_{1z}/\Delta_{0}$ with $|\Delta_{0}|\gg|\Omega_{1x}|,|\Omega_{1z}|$ . Here, $\Omega_{1x}$ and $\Omega_{1z}$ are the Rabi frequencies for the respective transitions shown in Fig.~\ref{fig:Schematics_Rb87}(b) and $\Delta_{0}$ is the single-photon detuning~\cite{qin2015three} of the excited state.
	
In general, the Rabi frequency $\Omega$ associated with the transition between two arbitrary states $|a\rangle,|b\rangle$ is given by $\Omega={\bf d}_{\rm eff}\cdot {\bf E}/\hbar$, where ${\bf E}$ is the electric field that drives the transition and ${\bf d}_{\rm eff}\equiv \langle b|\hat{\bf d}|a\rangle=-e\langle b|\hat{\bf r}|a\rangle$ is the effective dipole matrix element with $\hat{\bf r}=(\hat{x},\hat{y},\hat{z})$~\cite{wu2016realization}.

\item Type 2: Effective on-site SOC $M_r$ [shown again in Fig.~\ref{fig:Schematics_Rb87}(c)]
	
An additional SOC channel, which is on-site, is mediated by the excited state 5$^{2}$P$_{3/2}|1,-1\rangle$ and produced by two plane-wave Raman lasers with opposite wave vectors along the $x$ direction as
\begin{eqnarray}
{\bf E}_{2x}&\!=\!&E_{2x}e^{ik_{2}x+i\phi_{2x}}{\bf e}_{z},\\
{\bf E}'_{2x}&\!=\!&E'_{2x}e^{-ik_{2}x+i\phi'_{2x}}{\bf e}_{y},
\end{eqnarray}
where ${\bf e}_{z}$ and ${\bf e}_{y}$ are the directions of the polarization. As before, it can be shown with adiabatic elimination that the effective SOC amplitude between pseudospins $|\uparrow\rangle,|\downarrow\rangle$ is $M_{r}e^{i(2k_{2}x+\phi_{2})}$ with $M_r=-\hbar\Omega_{2x}\Omega'_{2x}/\Delta_r$, $|\Delta_{r}|\gg|\Omega_{2x}|,|\Omega'_{2x}|$ and $\phi_{2}=\phi_{2x}-\phi'_{2x}$. Here, $\Omega_{2x}$ and $\Omega'_{2x}$ are the respective Rabi frequencies and $\Delta_{r}$ is the single-photon detuning of the excited state [illustrated in Fig.~\ref{fig:Schematics_Rb87}(c)].
Experimentally, the ratio $k_2/k_1$ can be conveniently tuned by adjusting the propagating direction of ${\bf E}'_{2x}$ relative to ${\bf E}_{1x}$~\cite{zhou2022engineering}.
\end{enumerate}

\subsection{Laser-induced loss}\label{Supp_1.3}

Besides SOCs, optical driving can give rise to laser-induced loss (in addition to spontaneous decay) when the laser excites atoms to another state other than the ground-state pseudospins of interest. As shown in Fig.~\ref{fig:Schematics_Rb87}(d), a laser-induced loss term can be generated by coupling the hyperfine ground state $|\downarrow\rangle=|F=1,m_{F}=-1\rangle$ for 5$^2$S$_{1/2}$ to an excited state such as $|F\rq{}=2,m\rq{}_{F}=-2\rangle$ for 5$^2$P$_{1/2}$~\cite{li2019observation,lapp2019engineering,ren2022chiral}. We can write it as
\begin{equation}
\hat{H}_{\mathrm{loss}}=-i \gamma_{\downarrow}^{}\left|\downarrow\right\rangle\left\langle\downarrow\right|=i\gamma\sigma_z - i\gamma(\left|\uparrow\right\rangle\left\langle\uparrow\right|+\left|\downarrow\right\rangle\left\langle\downarrow\right|),\label{eq:loss}
\end{equation}
where $\gamma_{\downarrow}^{}=2\gamma$ is the effective decay rate. Here the overall loss term $-i\gamma(\left|\uparrow\right\rangle\left\langle\uparrow\right|+\left|\downarrow\right\rangle\left\langle\downarrow\right|)$
can be compensated by post-selection on $|\uparrow\rangle$ and $|\downarrow\rangle$ during experimental measurements~\cite{zhou2022engineering,shen2023proposal}. In our proposed experiment, $10^{5}$ atoms are  initialized in the system~\cite{liang2022dynamic}, and about $10^{4}$ atoms out of them should survive through our proposed short-time dynamics. Thus, we shall henceforth drop this overall loss term and keep only the $i\gamma\sigma_z$ term.

\subsection{Optical lattice potential}\label{Supp_1.4}

As a 2D system, an optical lattice potential along both the $x$ and $y$ directions is needed. In the $x$ direction, the optical lattice potential $V(x)$ is generated from the electric field ${\bf E}_{1x}$ [Eq.~(\ref{eq:E1x})], which also serves to generate the nearest-neighbor SOC. We have $V(x)=-V_{x}\cos^{2}(k_{1}x+\phi_{1-})$ with lattice depth $V_x$, lattice constant $a_x=\pi/k_{1}$, and the same phase offset $\phi_{1-}$ as in ${\bf E}_{1x}$ [Eq.~(\ref{eq:E1x})].

The optical lattice potential along the $y$ direction is generated by another electric field
\begin{eqnarray}
{\bf E}_{3y}\!=\!E_{3y}(e^{ik_{3}y\!+\!i\phi_{3y}}\!+\!e^{-ik_{3}y\!+\!i\phi'_{3y}}){\bf e}_{x}\!=\!E_{3y}e^{i\phi_{3+}}[e^{i(k_{3}y\!+\!\phi_{3-})}\!+\!e^{-i(k_{3}y\!+\!\phi_{3-})}]{\bf e}_{x}\!=\!2E_{3y}e^{i\phi_{3+}}\cos(k_{3}y\!+\!\phi_{3-}){\bf e}_{x},\label{eq:E3y}
\end{eqnarray} where ${\bf e}_{x}$ is the direction of the polarization, $\phi_{3\pm}=(\phi_{3y}\pm\phi'_{3y})/2$ with $\phi_{3y}$ and $\phi'_{3y}$ the phases of incident and reflected lights, respectively. It generates the $y$-lattice potential $V(y)=-V_{y}\cos^{2}(k_{3}y+\phi_{3-})$ with lattice depth $V_{y}$ and lattice constant $a_y=\pi/k_{3}$ along $y$ direction, and a phase offset $\phi_{3-}$ from ${\bf E}_{3y}$.

In all, the optical lattice potential then takes the form~\cite{wu2016realization}
\begin{equation}
V(x,y)=-V_{x}\cos^{2}(k_{1}x+\phi_{1-}) -V_{y}\cos^{2}(k_{3}y+\phi_{3-}).
\end{equation}
Its optical depths are given by $V_{x}=-\hbar|\Omega_{1x}|^{2}/\Delta_{1\sigma}$ ($\sigma=\uparrow,\downarrow$) and
$V_{y}=-\hbar|\Omega_{3y}|^{2}/\Delta_{2\sigma}$ with the detunings $\Delta_{1\sigma}$, $\Delta_{2\sigma}$ and Rabi frequencies $\Omega_{1x},\Omega_{3y}$ satisfying $|\Delta_{1\sigma}|\gg|\Omega_{1x}|$, and $|\Delta_{2\sigma}|\gg|\Omega_{3y}|$.

\subsection{Effective cold-atom model}\label{Supp_1.5}

Based on the above prescription for our proposed setup, the Hamiltonian density for our two-component ($\left|\uparrow\right\rangle$ and $\left|\downarrow\right\rangle$) atomic gas in a two-dimensional optical lattice (along $x$ and $y$ directions) is described by
\begin{eqnarray}
{\cal H}(x,y)
&\!=\!&\frac{\hat{\bf p}^2}{2m} \!+\! V(x,y) \!+\![{\rm A}(x)\sigma_{+} \!+\! {\rm h.c.}] \!+\! [{\rm B}(x)\sigma_{+} \!+\! {\rm h.c.}] \!+\! i\gamma\sigma_z \nonumber\\
&\!=\!& \frac{\hat{\bf p}^2}{2m} \!+\! V(x,y) \!+\! M_{0}\sin(k_{1}x)(e^{i\phi_0}\sigma_{+} \!+\! {\rm h.c.}) \!+\! M_{r}(e^{i\phi_r}\sigma_{+} \!+\! {\rm h.c.}) \!+\! i\gamma\sigma_z \nonumber\\
&\!=\!& \frac{\hat{\bf p}^2}{2m} \!+\! V(x,y) \!+\! M_{0}\sin(k_{1}x)(e^{i\phi_0}\sigma_{+} \!+\! {\rm h.c.}) \!+\! M_{r}(e^{i(2k_{2}x+\phi_2)}\sigma_{+} \!+\! {\rm h.c.}) \!+\! i\gamma\sigma_z \nonumber\\
&\!=\!& \frac{\hat{\bf p}^2}{2m} \!+\! V(x,y) \!+\! M_{0}\sin(k_{1}x)(e^{i\phi_0}\sigma_{+} \!+\! e^{-i\phi_0}\sigma_{-}) \!+\! M_{r}(e^{i(2k_{2}x+\phi_2)}\sigma_{+} \!+\! e^{-i(2k_{2}x+\phi_2)}\sigma_{-}) \!+\! i\gamma\sigma_z \nonumber\\
&\!=\!& \frac{\hat{\bf p}^2}{2m} \!+\! V(x,y) \!+\! \frac{M_{0}\sin(k_{1}x)}{2}[(e^{i\phi_0}\!+\!e^{-i\phi_0})\sigma_{x} \!+\! i(e^{i\phi_0}\!-\!e^{-i\phi_0})\sigma_{y}] \!+\! i\gamma\sigma_z \nonumber\\
&&~\!+\! \frac{M_{r}}{2}[(e^{i(2k_{2}x+\phi_2)}\!+\! e^{-i(2k_{2}x+\phi_2)})\sigma_{x} \!+\! i(e^{i(2k_{2}x+\phi_2)}\!-\! e^{-i(2k_{r}x+\phi_2)})\sigma_{y}] \nonumber\\
&\!=\!& \frac{\hat{\bf p}^2}{2m} \!+\! V(x,y) \!+\! M_{0}\sin(k_{1}x)(\cos\phi_0\sigma_{x} \!-\! \sin\phi_0\sigma_{y}) \!+\! M_{r}[\cos(2k_{2}x+\phi_2)\sigma_{x} \!-\! \sin(2k_{2}x+\phi_2)\sigma_{y}] \!+\! i\gamma\sigma_z \nonumber\\
&\!=\!& \frac{\hat{\bf p}^2}{2m} \!+\! V(x,y) \!+\! [M_{0}\sin(k_{1}x)\cos\phi_0 + M_{r}\cos(2k_{2}x+\phi_2) ]\sigma_{x} \!-\! [M_{0}\sin(k_{1}x)\sin\phi_0 \!+\! M_{r}\sin(2k_{r}x+\phi_2) ]\sigma_{y} \nonumber\\
&&~\!+\! i\gamma\sigma_{z},\label{eq:H_xy}
\end{eqnarray}
where the SOC terms are found to be ${\rm A}(x)=M_{0}\sin(k_{1}x)e^{i\phi_0}$ and ${\rm B}(x)=M_{r}e^{i\phi_r}$ with $\phi_r=2k_{2}x+\phi_{2}$. Here $\hat{\bf p}^2=\hat{p}_{x}^2+\hat{p}_{y}^2$ is the squared momentum operator, $V(x,y)= - V_{x}\cos^{2}(k_{1}x+\phi_{1-})- V_{y}\cos^{2}(k_{3}y+\phi_{3-})$ is the optical lattice potential and $\sigma_{\pm}=(\sigma_{x}\pm i\sigma_{y})/2$ with $\sigma_{j}$ ($j=x,y,z$) being Pauli matrices. The lattice constants along $x$ and $y$ directions are $a_x=\pi/k_{1}$ and $a_y=\pi/k_{3}$, respectively.

\subsubsection{Tight-binding model}\label{Supp_1.5.1}

To obtain the effective tight-binding model from the above Hamiltonian density, we expand the field operator $\hat{\psi}_{\sigma}(x,y)=\sum_{j_x,j_y}W_0(x-a_{x}j_{x},y-a_{y}j_{y})\hat{c}_{j_x,j_y,\sigma}^{}$, where $\sigma=\uparrow,\downarrow$, $a_x$ and $a_y$ are the lattice constants along $x$ and $y$ directions, respectively, $W_{0}(x,y)$ is a maximally localized Wannier function~\cite{marzari1997maximally,lee2014lattice} of the lowest-band ($n=0$), $j_x$ and $j_y$ are the indexes of lattice sites along $x$ and $y$ directions, respectively. In this way, the second-quantized single-particle Hamiltonian can be reduced to the tight-banding model, with the following parameters:

The nearest-neighbor-hopping terms are
\begin{eqnarray}
t_x&=&-\int dx\int dyW_{0}^{*}(x,y)\left[ \frac{\hat{p}_x^2}{2m} \!-\! V_{x}\cos^{2}(k_{1}x+\phi_{1-}) \right]W_{0}(x-a_x,y),\\
t_y&=&-\int dx\int dyW_{0}^{*}(x,y)\left[ \frac{\hat{p}_y^2}{2m} \!-\! V_{y}\cos^{2}(k_{3}y+\phi_{3-}) \right]W_{0}(x,y-a_y).
\end{eqnarray}
The on-site spin-flip-hopping terms are
\begin{eqnarray}
t_{\uparrow\downarrow}^{j_x}&=&\int dx\int dyW_{0}^{*}(x-x_{j_x},y)M_{r}e^{i(2k_{2}x+\phi_2)}W_{0}(x-x_{j_x},y)\equiv e^{i\phi_{rr}j_x}\Omega_{r},\\
t_{\downarrow\uparrow}^{j_x}&=&\int dx\int dyW_{0}^{*}(x-x_{j_x},y)M_{r}e^{-i(2k_{2}x+\phi_2)}W_{0}(x-x_{j_x},y)\equiv e^{-i\phi_{rr}j_x}\Omega_{r},
\end{eqnarray}
where we use $x_{j_x}=a_{x}j_{x}$, $\phi_{rr}=2k_{2}a_{x}=2\pi k_{2}/k_{1}$, and $\Omega_{r}=\int dx\int dyW_{0}^{*}(x,y)M_{r}e^{i(2k_{2}x+\phi_2)}W_{0}(x,y)$, $M_r=-\hbar \Omega_{2x}\Omega'_{2x}/\Delta_r$.

The nearest-neighbor spin-flip terms are
\begin{eqnarray}
t_{\uparrow\downarrow}^{j_x,j_x+1}&=&\int dx\int dyW_{0}^{*}(x-x_{j_x},y)[M_{0}\sin(k_{1}x)e^{i\phi_0}]W_{0}(x-x_{j_x+1},y) \nonumber\\
&=&\int dx\int dyW_{0}^{*}(x,y)[M_{0}\sin(k_{1}(x+x_{j_x}))e^{i\phi_0}]W_{0}(x-a_{x},y) \nonumber\\
&=&e^{i\phi_0}(-1)^{j_x}\int dx\int dyW_{0}^{*}(x,y)[M_{0}\sin(k_{1}x)]W_{0}(x-a_{x},y)\equiv e^{i\phi_{0}}(-1)^{j_x}\Omega_{0},
\end{eqnarray}
\begin{eqnarray}
t_{\uparrow\downarrow}^{j_x,j_x-1}&=&\int dx\int dyW_{0}^{*}(x-x_{j_x},y)[M_{0}\sin(k_{1}x)e^{i\phi_0}]W_{0}(x-x_{j_x-1},y) \nonumber\\
&=&\int dx\int dyW_{0}^{*}(x,y)[M_{0}\sin(k_{1}(x+x_{j_x}))e^{i\phi_0}]W_{0}(x+a_{x},y) \nonumber\\
&=&e^{i\phi_0}(-1)^{j_x}\int dx\int dyW_{0}^{*}(x,y)[M_{0}\sin(k_{1}x)]W_{0}(x+a_{x},y) \nonumber\\
&=&e^{i\phi_{0}}(-1)^{j_x}\int dx\int dyW_{0}^{*}(x-a_x,y)[M_{0}\sin(k_{1}(x-a_x))]W_{0}(x,y) \nonumber\\
&=&-e^{i\phi_{0}}(-1)^{j_x}\int dx\int dyW_{0}^{*}(x-a_x,y)[M_{0}\sin(k_{1}x)]W_{0}(x,y)\equiv - e^{i\phi_{0}}(-1)^{j_x}\Omega_{0}=-t_{\uparrow\downarrow}^{j_x,j_x+1},
\end{eqnarray}
\begin{eqnarray}
t_{\downarrow\uparrow}^{j_x+1,j_x}&=&\int dxW_{0}^{*}(x-x_{j_x+1},y)[M_{0}\sin(k_{1}x)e^{i\phi_0}]W_{0}(x-x_{j_x},y) \nonumber\\
&=&\int dx\int dyW_{0}^{*}(x,y)[M_{0}\sin(k_{1}(x+x_{j_x+1}))e^{i\phi_0}]W_{0}(x+a_{x},y) \nonumber\\
&=&e^{i\phi_{0}}(-1)^{j_x+1}\int dx\int dyW_{0}^{*}(x,y)[M_{0}\sin(k_{1}x)]W_{0}(x+a_{x},y) \nonumber\\
&=&-e^{i\phi_{0}}(-1)^{j_x}\int dx\int dyW_{0}^{*}(x-a_x,y)[M_{0}\sin(k_{1}(x-a_x))]W_{0}(x,y) \nonumber\\
&=&e^{i\phi_{0}}(-1)^{j_x}\int dx\int dyW_{0}^{*}(x-a_x,y)[M_{0}\sin(k_{1}x)]W_{0}(x,y)\equiv e^{i\phi_{0}}(-1)^{j_x}\Omega_{0}=t_{\uparrow\downarrow}^{j_x,j_x+1},
\end{eqnarray} where $\Omega_{0}=\int dx\int dyW_{0}^{*}(x,y)[M_{0}\sin(k_{1}x)]W_{0}(x-a_x,y)=\int dx\int dyW_{0}^{*}(x-a_x,y)[M_{0}\sin(k_{1}x)]W_{0}(x,y)$, $M_0=-\hbar \Omega_{1x}\Omega_{1z}/\Delta_0$, and we have used the identities
\begin{eqnarray}
\sin(k_{1}(x+x_{j_x}))&\!=\!&\sin(k_{1}(x+a_{x}j_{x}))\!=\!\sin(k_{1}x+\pi j_{x})\!=\!(-1)^{j_x}\sin(k_{1}x),\\
\sin(k_{1}(x+x_{j_x+1}))&\!=\!&\sin(k_{1}(x+a_{x}(j_{x}+1)))\!=\!\sin(k_{1}x+\pi (j_{x}+1))\!=\!(-1)^{j_x+1}\sin(k_{1}x)\!=\!-(-1)^{j_x}\sin(k_{1}x),\\
\sin(k_{1}(x-a_{x}))&\!=\!&\sin(k_{1}x-\pi)\!=\!-\sin(k_{1}x).
\end{eqnarray}

The above overlap integrals allow us to write down the effective Hamiltonian in the following tight-binding form:
\begin{eqnarray}
\hat{H}_{\rm tb}&=&-t_{x}\sum_{j_x,j_y}(\hat{c}_{j_x,j_y,\uparrow}^{\dagger}\hat{c}_{j_{x}+1,j_y,\uparrow}^{} + \hat{c}_{j_{x},j_y,\downarrow}^{\dagger}\hat{c}_{j_{x}+1,j_y,\downarrow}^{} + {\rm h.c.})
	-t_{y}\sum_{j_x,j_y}(\hat{c}_{j_x,j_y,\uparrow}^{\dagger}\hat{c}_{j_{x},j_y+1,\uparrow}^{} + \hat{c}_{j_{x},j_y,\downarrow}^{\dagger}\hat{c}_{j_{x},j_y+1,\downarrow}^{} + {\rm h.c.}) \nonumber\\
&&~+\Omega_{0}\sum_{j_x,j_y}[e^{i\phi_0}(-1)^{j_x}(\hat{c}_{j_x,j_y,\uparrow}^{\dagger}\hat{c}_{j_{x}+1,j_y,\downarrow}^{} + \hat{c}_{j_x+1,j_y,\uparrow}^{\dagger}\hat{c}_{j_{x},j_y,\downarrow}^{}) + {\rm h.c.}] \nonumber\\
&&~+\Omega_{r}\sum_{j_x,j_y}(e^{i\phi_{rr}j_x}\hat{c}_{j_x,j_y,\uparrow}^{\dagger}\hat{c}_{j_{x},j_y,\downarrow}^{} + {\rm h.c.}) + i\gamma\sum_{j_x,j_y}(\hat{c}_{j_x,j_y,\uparrow}^{\dagger}\hat{c}_{j_{x},j_y,\uparrow}^{} - \hat{c}_{j_x,j_y,\downarrow}^{\dagger}\hat{c}_{j_{x},j_y,\downarrow}^{}).\label{eq:H_tb_1}
\end{eqnarray}

\noindent Redefining the spin-down operator $\hat{c}_{j_x,j_y,\downarrow}^{}\rightarrow e^{i\pi (x_{j_x}/a_{x}+y_{j_y}/a_{y})}\hat{c}_{j_x,j_y,\downarrow}^{}=(-1)^{j_x+j_y}\hat{c}_{j_x,j_y,\downarrow}^{}$~\cite{wu2016realization,zhou2022engineering} with $x_{j_x}=a_{x}j_{x}$, $y_{j_y}=a_{y}j_{y}$, $j_{y}=2n_{y}$, $n_{y}\in\mathbb{N}_{+}$, $\mathbb{N}_{+}$ for positive integers, and $\phi_{rr}=\pi$ (i.e., $2k_{2}=k_{1}$), i.e., performing a gauge transformation, we recast the Hamiltonian into
\begin{eqnarray}
\hat{H}_{\rm tb}&=&-t_{x}\sum_{j_x,j_y}(\hat{c}_{j_x,j_y,\uparrow}^{\dagger}\hat{c}_{j_{x}+1,j_y,\uparrow}^{} - \hat{c}_{j_{x},j_y,\downarrow}^{\dagger}\hat{c}_{j_{x}+1,j_y,\downarrow}^{} + {\rm h.c.})
	-t_{y}\sum_{j_x,j_y}(\hat{c}_{j_x,j_y,\uparrow}^{\dagger}\hat{c}_{j_{x},j_y+1,\uparrow}^{} - \hat{c}_{j_{x},j_y,\downarrow}^{\dagger}\hat{c}_{j_{x},j_y+1,\downarrow}^{} + {\rm h.c.}) \nonumber\\
&&~+\Omega_{0}\sum_{j_x,j_y}[-e^{i\phi_0}(\hat{c}_{j_x,j_y,\uparrow}^{\dagger}\hat{c}_{j_{x}+1,j_y,\downarrow}^{} - \hat{c}_{j_x+1,j_y,\uparrow}^{\dagger}\hat{c}_{j_{x},j_y,\downarrow}^{}) + {\rm h.c.}] \nonumber\\
&&~+\Omega_{r}\sum_{j_x,j_y}(\hat{c}_{j_x,j_y,\uparrow}^{\dagger}\hat{c}_{j_{x},j_y,\downarrow}^{} + {\rm h.c.}) + i\gamma\sum_{j_x,j_y}(\hat{c}_{j_x,j_y,\uparrow}^{\dagger}\hat{c}_{j_{x},j_y,\uparrow}^{} - \hat{c}_{j_x,j_y,\downarrow}^{\dagger}\hat{c}_{j_{x},j_y,\downarrow}^{}).\label{eq:H_tb_2}
\end{eqnarray}
For the case $j_{y}=2n_{y}-1$, one can find $\Omega_{0}\to-\Omega_{0}$ and $\Omega_{r}\to-\Omega_{r}$ in Eq.~\eqref{eq:H_tb_2}.

\subsubsection{Effective Hamiltonian in momentum space}\label{Supp_1.5.2}

Performing a 2D Fourier transformation, one obtains
\begin{eqnarray}
\hat{c}_{j_x,j_y,\sigma}^{}&=&\sum_{k_x,k_y}e^{i(k_{x}a_{x}j_{x}+k_{y}a_{y}j_{y})}\hat{c}_{k_x,k_y,\sigma}^{},\\
\hat{c}_{j_x,j_y,\sigma}^{\dagger}&=&\sum_{k_x,k_y}e^{-i(k_{x}a_{x}j_{x}+k_{y}a_{y}j_{y})}\hat{c}_{k_x,k_y,\sigma}^{\dagger}.
\end{eqnarray}
Therefore, our effective Hamiltonian in momentum space is given by
\begin{eqnarray}
\hat{H}_{\rm eff}&\!=\!&\sum_{k_x,k_y}\left[-t_{x}e^{ik_xa_x}\hat{c}_{k_x,k_y,\uparrow}^{\dagger}\hat{c}_{k_x,k_y,\uparrow}^{} + t_{x}e^{ik_xa_x}\hat{c}_{k_x,k_y,\downarrow}^{\dagger}\hat{c}_{k_x,k_y,\downarrow}^{} - t_{y}e^{ik_ya_y}\hat{c}_{k_x,k_y,\uparrow}^{\dagger}\hat{c}_{k_x,k_y,\uparrow}^{} + t_{y}e^{ik_ya_y}\hat{c}_{k_x,k_y,\downarrow}^{\dagger}\hat{c}_{k_x,k_y,\downarrow}^{} \right.\nonumber\\
&&\left.~\!-\!\Omega_{0}e^{i\phi_{0}}(e^{ik_xa_x}\!-\!e^{-ik_xa_x})\hat{c}_{k_x,k_y,\uparrow}^{\dagger}\hat{c}_{k_x,k_y,\downarrow}^{} \!+\! \Omega_{r}\hat{c}_{k_x,k_y,\uparrow}^{\dagger}\hat{c}_{k_x,k_y,\downarrow}^{} \right] \!+\! {\rm h.c.} \!+\! i\gamma\sum_{k_x,k_y}(\hat{c}_{k_x,k_y,\uparrow}^{\dagger}\hat{c}_{k_x,k_y,\uparrow}^{} \!-\! \hat{c}_{k_x,k_y,\downarrow}^{\dagger}\hat{c}_{k_x,k_y,\downarrow}^{}) \nonumber\\
&=&\sum_{k_x,k_y}\left[(-t_{x}e^{ik_xa_x} - t_{y}e^{ik_ya_y} + {\rm h.c.} + i\gamma)\hat{c}_{k_x,k_y,\uparrow}^{\dagger}\hat{c}_{k_x,k_y,\uparrow}^{} + (t_{x}e^{ik_xa_x} + t_{y}e^{ik_ya_y} + {\rm h.c.} - i\gamma )\hat{c}_{k_x,k_y,\downarrow}^{\dagger}\hat{c}_{k_x,k_y,\downarrow}^{} \right.\nonumber\\
&&\left.~+ [\Omega_{r} - i2\Omega_{0}e^{i\phi_{0}}\sin(k_xa_x) ]\hat{c}_{k_x,k_y,\uparrow}^{\dagger}\hat{c}_{k_x,k_y,\downarrow}^{} + [\Omega_{r} + i2\Omega_{0}e^{-i\phi_{0}}\sin(k_xa_x) ]\hat{c}_{k_x,k_y,\downarrow}^{\dagger}\hat{c}_{k_x,k_y,\uparrow}^{} \right] \nonumber\\
&\!=\!&\sum_{k_x,k_y}(\hat{c}_{k_x,k_y,\uparrow}^{\dagger}, \hat{c}_{k_x,k_y,\downarrow}^{\dagger}) \nonumber\\
&&~\times\!
\begin{pmatrix}
\!-t_{x}(e^{ik_xa_x} \!+\! e^{-ik_xa_x}) \!-\! t_{y}(e^{ik_ya_y} \!+\! e^{-ik_ya_y}) \!+\! i\gamma & \Omega_{r} \!-\! i2\Omega_{0}e^{i\phi_{0}}\sin(k_xa_x) \\
\Omega_{r} \!+\! i2\Omega_{0}e^{-i\phi_{0}}\sin(k_xa_x) & t_{x}(e^{ik_xa_x} \!+\! e^{-ik_xa_x}) \!+\! t_{y}(e^{ik_ya_y} \!+\! e^{-ik_ya_y}) \!-\! i\gamma\!
\end{pmatrix}\!\!\begin{pmatrix}
\!\hat{c}_{k_x,k_y,\uparrow}^{}\! \\
\!\hat{c}_{k_x,k_y,\downarrow}^{}\!
\end{pmatrix}\! \nonumber\\
&\!=\!&\sum_{k_x,k_y}(\hat{c}_{k_x,k_y,\uparrow}^{\dagger}, \hat{c}_{k_x,k_y,\downarrow}^{\dagger}) \nonumber\\
&&~\times\!\begin{pmatrix}
\!-2t_{x}\cos (k_xa_x) \!-\! 2t_{y}\cos(k_ya_y) \!+\! i\gamma & \Omega_{r} \!+\! 2\Omega_{0}\sin\phi_{0}\sin (k_xa_x) \!-\! i2\Omega_{0}\cos\phi_{0}\sin(k_xa_x) \\
\Omega_{r} \!+\! 2\Omega_{0}\sin\phi_{0}\sin (k_xa_x) \!+\! i2\Omega_{0}\cos\phi_{0}\sin(k_xa_x) & 2t_{x}\cos(k_xa_x) \!+\! 2t_{y}\cos(k_ya_y) \!-\! i\gamma\!
\end{pmatrix}\!\!\begin{pmatrix}
\!\hat{c}_{k_x,k_y,\uparrow}^{}\! \\
\!\hat{c}_{k_x,k_y,\downarrow}^{}\!
\end{pmatrix}\! \nonumber\\
&\!=\!&\sum_{k_x,k_y}(\hat{c}_{k_x,k_y,\uparrow}^{\dagger}, \hat{c}_{k_x,k_y,\downarrow}^{\dagger})
\!\begin{pmatrix}
\!i\gamma \!-\! 2t_{x}\cos (k_xa_x) \!-\! 2t_{y}\cos (k_ya_y) & \Omega_{r} \!-\! i2\Omega_{0}e^{i\phi_{0}}\sin (k_xa_x) \\
\Omega_{r} + i2\Omega_{0}e^{-i\phi_{0}}\sin (k_xa_x) & \!-\!i\gamma \!+\! 2t_{x}\cos (k_xa_x) \!+\! 2t_{y}\cos (k_ya_y) \!
\end{pmatrix}\!\!\begin{pmatrix}
\!\hat{c}_{k_x,k_y,\uparrow}^{}\! \\
\!\hat{c}_{k_x,k_y,\downarrow}^{}\!
\end{pmatrix}\!, \label{eq:H_kx_ky_0}
\end{eqnarray} where we have used
\begin{eqnarray}
\sum_{k'_x,k'_y}\sum_{k_x,k_y}e^{i(k_x-k'_x)a_xj_{x}}e^{i(k_y-k'_y)a_yj_{y}} = \sum_{k'_x,k'_y}\delta_{k_x,k'_x}\delta_{k_y,k'_y}\rightarrow\sum_{k_x,k_y}.
\end{eqnarray}

Furthermore, in the basis $\hat{\Psi}_{k_x,k_y}=(\hat{c}_{k_x,k_y,\uparrow}^{},\hat{c}_{k_x,k_y,\downarrow}^{})^{T}$, our model Hamiltonian in momentum space can be written as
\begin{eqnarray}
\hat{\cal H}_{\rm eff}(k_x,k_y)&=&
\begin{pmatrix}
i\gamma - 2t_{x}\cos (k_xa_x) - 2t_{y}\cos (k_ya_y) & \Omega_{r} - i2\Omega_{0}e^{i\phi_{0}}\sin (k_xa_x) \nonumber\\
\Omega_{r} + i2\Omega_{0}e^{-i\phi_{0}}\sin (k_xa_x) & -i\gamma + 2t_{x}\cos (k_xa_x) + 2t_{y}\cos (k_ya_y)
\end{pmatrix} \label{eq:H_kx_ky_00}\\
&=&
\begin{pmatrix}
i\gamma - t_{x}(z+z^{-1}) - 2t_{y}\cos (k_ya_y) & \Omega_{r} - \Omega_{0}e^{i\phi_{0}}(z-z^{-1}) \\
\Omega_{r} + \Omega_{0}e^{-i\phi_{0}}(z-z^{-1}) & -i\gamma + t_{x}(z+z^{-1}) + 2t_{y}\cos (k_ya_y)
\end{pmatrix} \nonumber\\
&=&d_x\sigma_x + d_y\sigma_y + d_z\sigma_z, \label{eq:H_kx_ky}
\end{eqnarray} where $z=e^{ik_xa_x}$, and we have used $\cos(k_xa_x)=(e^{ik_xa_x}+e^{-ik_xa_x})/2$, $\sin (k_xa_x)=(e^{ik_xa_x}-e^{-ik_xa_x})/(2i)$,
\begin{equation}\label{eq:Hdxyz}
\begin{aligned}
d_x &= \Omega_{r} + 2\Omega_{0}\sin\phi_{0}\sin (k_xa_x)= \Omega_{r} - i\Omega_{0}\sin\phi_{0}(z-z^{-1}), \\
d_y &= 2\Omega_{0}\cos\phi_{0}\sin (k_xa_x) = -i\Omega_{0}\cos\phi_{0}(z-z^{-1}), \\
d_z &= i\gamma - 2t_{x}\cos (k_xa_x) - 2t_{y}\cos (k_ya_y) = i\gamma - t_{x}(z+z^{-1}) - 2t_{y}\cos (k_ya_y).
\end{aligned}
\end{equation}
Therefore, the eigenenergies take the form of
\begin{eqnarray}
E^{2}(k_x,k_y)&=&d_x^2 + d_y^2 + d_z^2 \nonumber\\
&=&[\Omega_{r} - i\Omega_{0}\sin\phi_{0}(z-z^{-1})]^{2} + [-i\Omega_{0}\cos\phi_{0}(z-z^{-1})]^{2} + [i\gamma - t_{x}(z+z^{-1}) - 2t_{y}\cos (k_ya_y)]^{2}  \nonumber\\
&=&t_{x}^{2}(z + z^{-1})^{2} - \Omega_{0}^{2}(z-z^{-1})^{2} - i2\Omega_{r}\Omega_{0}\sin\phi_{0}(z-z^{-1}) - 2t_{x}[i\gamma - 2t_{y}\cos (k_ya_y)](z+z^{-1}) \nonumber\\
&&+\Omega_{r}^{2} - \gamma^{2} + 4t_{y}^{2}\cos^{2} (k_ya_y) - i4\gamma t_{y}\cos (k_ya_y).
\end{eqnarray}

By enforcing $t_{x}=\Omega_{0}$ (through adjustment of the physical parameters) in order to obtain our desired model, we have
\begin{eqnarray}
E^{2}(k_x,k_y)&=&-2t_{x}[i\Omega_{r}\sin\phi_{0} + i\gamma - 2t_{y}\cos (k_{y}a_y)]z - 2t_{x}[-i\Omega_{r}\sin\phi_{0} + i\gamma - 2t_{y}\cos (k_{y}a_y)]z^{-1} \nonumber\\
&&+ \Omega_{r}^{2} - \gamma^{2} + 4t_{x}^{2} + 4t_{y}^{2}\cos^{2} (k_ya_y) - i4\gamma t_{y}\cos (k_ya_y) \nonumber\\
&=&t_{+}z + t_{-}z^{-1} + t_{0},
\end{eqnarray} where
\begin{eqnarray}
t_{+}&=&-2t_{x}[i\Omega_{r}\sin\phi_{0} + i\gamma - 2t_{y}\cos (k_ya_y)]= -i2t_{x}[\Omega_{r}\sin\phi_{0} + \gamma + i2t_{y}\cos (k_ya_y)],\\
t_{-}&=&- 2t_{x}[-i\Omega_{r}\sin\phi_{0} + i\gamma - 2t_{y}\cos (k_ya_y)]= - i2t_{x}[-\Omega_{r}\sin\phi_{0} + \gamma + i2t_{y}\cos (k_ya_y)],\\
t_{0}&=&\Omega_{r}^{2} - \gamma^{2} + 4t_{x}^{2} + 4t_{y}^{2}\cos^{2} (k_ya_y) - i4\gamma t_{y}\cos (k_ya_y).
\end{eqnarray}

If we define the dimensionless wave numbers as $k_x\leftarrow k_xa_x$ and $k_y\leftarrow k_ya_y$ for simplicity, we can rewrite Eq.~\eqref{eq:Hdxyz} as
\begin{equation}
\begin{aligned}
&d_x = \Omega_{r} + 2\Omega_{0}\sin\phi_{0}\sin k_x,\\
&d_y=2\Omega_{0}\cos\phi_{0}\sin k_x,\\
&d_z = i\gamma - 2t_{x}\cos k_x - 2t_{y}\cos k_y,
\end{aligned}
\end{equation} which is consistent with Eq.~\eqref{eq:d} of the main text.


\section{Band singularities from non-Hermitian pumping}\label{Supp_2}

\subsection{Complex deformation from non-Hermitian pumping}\label{Supp_2.1}

We consider cylindrical boundary conditions for our Hamiltonian of Eq.~\eqref{eq:H_kx_ky} with OBCs in the $x$ direction and PBCs in the $y$ direction. Here we set the dimensionless crystal momenta as $k_x\leftarrow k_xa_x$ and $k_y\leftarrow k_ya_y$ for simplicity, and rewrite Eq.~\eqref{eq:H_kx_ky_00} as follows:
\begin{equation}\label{eq:H_kappa}
\hat{\cal H}_{\rm eff}(k_x,k_y)=
\begin{pmatrix}
i\gamma - 2t_{x}\cos k_x - 2t_{y}\cos k_y & \Omega_{r} - i2\Omega_{0}e^{i\phi_{0}}\sin k_x \\
\Omega_{r} + i2\Omega_{0}e^{-i\phi_{0}}\sin k_x & -i\gamma + 2t_{x}\cos k_x + 2t_{y}\cos k_y
\end{pmatrix}.
\end{equation}
As discussed in the main text, non-Hermitian pumping in the $x$ direction under $x$ OBCs leads to the complex deformation $k_x\rightarrow k_x+i\kappa_x({\bf k})$~\cite{yao2018edge,okuma2020topological,li2020critical}
\begin{eqnarray}
\kappa_{x}(k_y) = \ln\sqrt{\frac{t_{+}}{t_{-}}} = \ln\sqrt{\frac{\gamma + \Omega_{r}\sin\phi_{0} + i2t_{y}\cos k_y}{\gamma - \Omega_{r}\sin\phi_{0} + i2t_{y}\cos k_y}}.
\end{eqnarray}
\begin{figure}[h]
\centering
\includegraphics[width=0.35\linewidth]{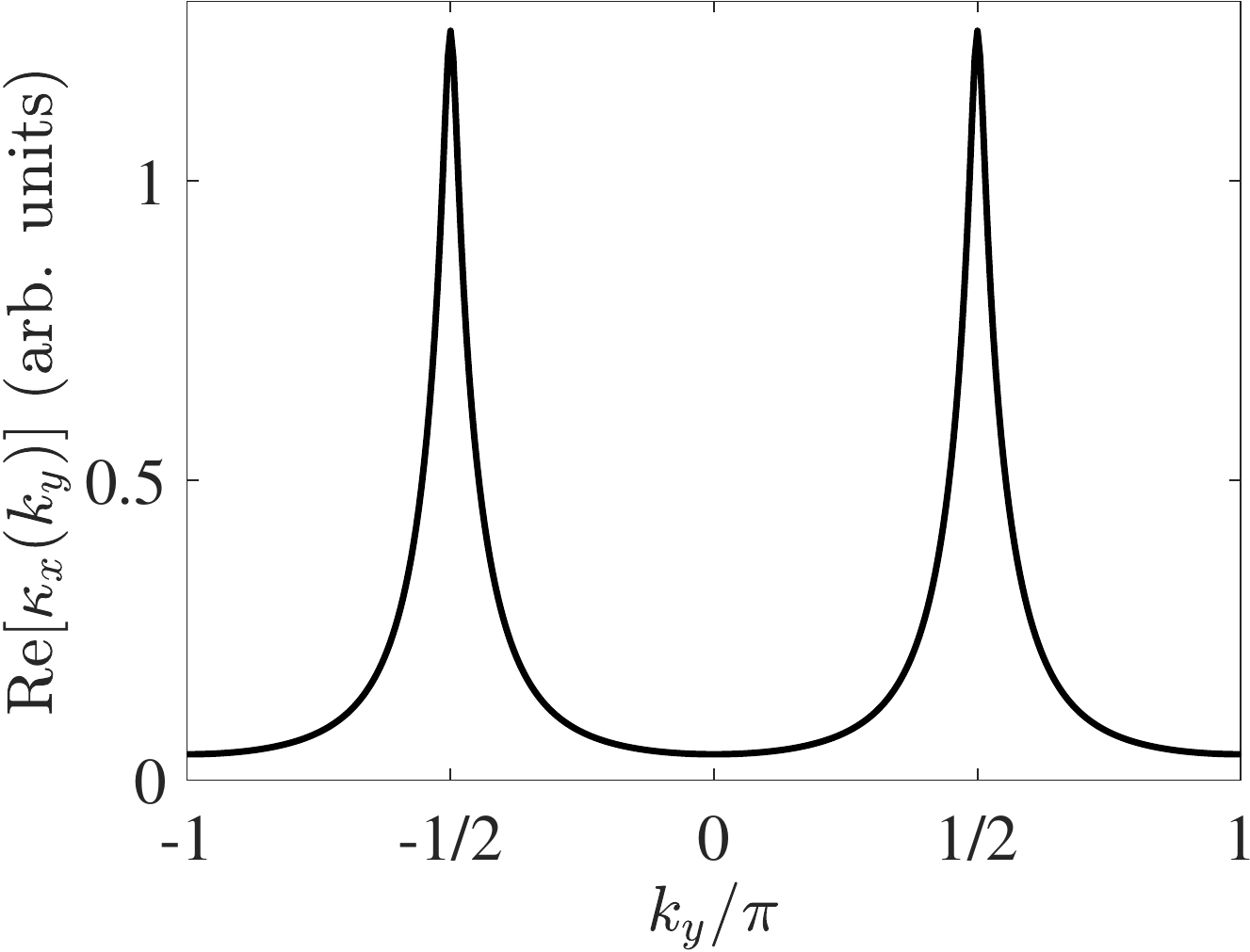}
\caption{Plot of the real part of $\kappa_{x}(k_{y})$. We set the parameters as $\phi_{0}=\pi/4$, $\Omega_{0}/t_{x}=1$, $t_{y}/t_{x}=3$, $\Omega_{r}/t_{x}=2$, $\gamma/t_{x}=1.2$, with $t_{x}$ as the energy unit. Divergences appear at $k_{y}=\pm\frac{\pi}{2}$, where $\text{Im}[\kappa_x(k_{y})]$ vanishes. Without loss of generality, these parameters are in arbitrary units (arb. units).}\label{fig:fky}
\end{figure}
As shown in Fig.~\ref{fig:fky}, the divergences appear at $k_{y}=\pm\frac{\pi}{2}$, where $\text{Im}[\kappa_x(k_{y})]$ vanishes. Under the complex deformation $k_x\rightarrow k_x+i\kappa_x({\bf k})$, with parameters specialized to $t_{x}=\Omega_{0}$, the eigenenergies of Eq.~\eqref{eq:H_kappa} take the form
\begin{eqnarray}
&&E^{2}(\bar{k}_x,k_y)=E^{2}(k_x+i\kappa_x(k_y) ,k_y)=\Omega_{r}^{2}-\gamma^{2}+4 t_x^{2}+4t_y^{2}\cos^{2}{k_y}-4 i\gamma {t_y}\cos{k_y} \nonumber\\
&&\!-4 i \gamma t_x\cos\!\left(\!\!k_x\!+\!i\ln\sqrt{\frac{\gamma \!+\! \Omega_{r}\sin\phi_{0} \!+\! i2t_{y}\cos k_y}{\gamma \!-\! \Omega_{r}\sin\phi_{0} \!+\! i2t_{y}\cos k_y}}\right)\!+\!4t_x t_y\cos\!\left(\!\!k_x\!+\!i\ln\sqrt{\frac{\gamma \!+\! \Omega_{r}\sin\phi_{0} \!+\! i2t_{y}\cos k_y}{\gamma \!-\! \Omega_{r}\sin\phi_{0} \!+\! i2t_{y}\cos k_y}}\!+\!k_y\!\!\right) \nonumber\\
&&\!+4t_x t_y\cos\!\left(\!\!k_x\!+\!i\ln\sqrt{\frac{\gamma \!+\! \Omega_{r}\sin\phi_{0} \!+\! i2t_{y}\cos k_y}{\gamma \!-\! \Omega_{r}\sin\phi_{0} \!+\! i2t_{y}\cos k_y}}\!-\!k_y\!\!\right)\!+\!4\sin\phi_{0}t_x\Omega_r\sin\!\left(\!\!k_x\!+\!i\ln\sqrt{\frac{\gamma \!+\! \Omega_{r}\sin\phi_{0} \!+\! i2t_{y}\cos k_y}{\gamma \!-\! \Omega_{r}\sin\phi_{0} \!+\! i2t_{y}\cos k_y}}\right)\!, \label{eq:eigenenergies_E}
\end{eqnarray} where we use $\cos(k_x+i\kappa)\cos k_y=\frac{1}{2}[\cos(k_x+i\kappa+k_y)+\cos(k_x+i\kappa-k_y)]$.
To express this as a Hatano-Nelson model along the $x$-direction with $k_y$-valued coefficients, we make use of
\begin{equation}
\begin{aligned}
&\sin\!\left(\!\!k_{x}\!+\!i\ln\sqrt{\frac{\gamma + \Omega_{r}\sin\phi_{0} + i2t_{y}\cos k_y}{\gamma - \Omega_{r}\sin\phi_{0} + i2t_{y}\cos k_y}}\right)
\!=\!\frac{e^{ik_x}}{2i\sqrt{\frac{\gamma + \Omega_{r}\sin\phi_{0} + i2t_{y}\cos k_y}{\gamma - \Omega_{r}\sin\phi_{0} + i2t_{y}\cos k_y}}} \!-\! \frac{e^{-ik_x}}{2i}\sqrt{\frac{\gamma + \Omega_{r}\sin\phi_{0} + i2t_{y}\cos k_y}{\gamma - \Omega_{r}\sin\phi_{0} + i2t_{y}\cos k_y}}, \\
&\cos\!\left(\!\!k_{x}\!+\!i\ln\sqrt{\frac{\gamma + \Omega_{r}\sin\phi_{0} + i2t_{y}\cos k_y}{\gamma - \Omega_{r}\sin\phi_{0} + i2t_{y}\cos k_y}}\right)
\!=\!\frac{e^{ik_x}}{2\sqrt{\frac{\gamma + \Omega_{r}\sin\phi_{0} + i2t_{y}\cos k_y}{\gamma - \Omega_{r}\sin\phi_{0} + i2t_{y}\cos k_y}}}\!+\!\frac{e^{-ik_x}}{2}\sqrt{\frac{\gamma + \Omega_{r}\sin\phi_{0} + i2t_{y}\cos k_y}{\gamma - \Omega_{r}\sin\phi_{0} + i2t_{y}\cos k_y}}.
\end{aligned}
\end{equation}
With that, Eq.~\eqref{eq:eigenenergies_E} simplifies to
\begin{equation}\label{dis}
E^{2}(\bar{k}_x,k_y)=\sqrt{t_{+}t_{-}}(e^{ik_{x}}+e^{-ik_{x}})+t_{0},
\end{equation} with
\begin{eqnarray}
t_{+}&=& -2it_{x}\left(\gamma + \Omega_{r}\sin\phi_{0} + 2it_{y}\cos k_y \right),\\
t_{-}&=& - 2it_{x}\left(\gamma - \Omega_{r}\sin\phi_{0} + 2it_{y}\cos k_y \right),\\
t_{0}&=&\Omega_{r}^{2} - \gamma^{2} + 4t_{x}^{2} + 4t_{y}^{2}\cos^{2}k_y - i4\gamma t_{y}\cos k_{y}.
\end{eqnarray}
This energy dispersion is manifestly even along the $\hat{k}_{x}$ and $\hat{k}_{y}$ directions, as required by our response kink setup in the text.

With $E(\bar{\bf k})$ being that of the lower occupied band, we have
\begin{equation}\label{dis_E}
E(\bar{k}_x,k_y)=-\sqrt{\sqrt{t_{+}t_{-}}(e^{ik_{x}}+e^{-ik_{x}})+t_{0}}.
\end{equation} Therefore, $E(\bar{k}_x,k_y)$ is an even function of $k_x$, so that $\partial E(\bar{k}_x,k_y)/\partial k_{x}$ is an odd funciton of $k_x$ as follows:
\begin{equation}
\frac{\partial E(\bar{k}_x,k_y)}{\partial k_{x}}
=\frac{\sqrt{t_{+}t_{-}}\sin k_x}{\sqrt{t_0+2\sqrt{t_{+}t_{-}}\cos k_x}}.
\end{equation}
Furthermore, we have
\begin{equation}
\frac{\partial E(\overline{\mathbf{k}}_{+}(t))}{\partial k_{x}}+\frac{\partial E(\overline{\mathbf{k}}_{-}(t))}{\partial k_{x}}
=0,
\end{equation} where $\bar{\bf k}_{\pm}(t)={\bf k}_{\pm}(t) + i\kappa_{x}({\bf k}_{\pm}(t))$, ${\mathbf{k}}_{\pm}(t)=\hbar^{-1}\int_{0}^{t}{\bf F}_{\pm}dt^{\prime}$ with ${\bf F}_{\pm}=F_{y}(t){\bf e}_{y}\pm F_{x}{\bf e}_{x}$.

\subsection{Discontinuities of the Berry curvature}\label{Supp_2.2}

To compute the biorthogonal Berry curvature in a non-Hermitian system, with or without non-Hermitian pumping, note that the left (bra) and right (ket) eigenstates are not identical:
\begin{eqnarray}
\hat{\cal H}_{\rm eff}|\psi_{R}^{}\rangle&=&E|\psi_{R}^{}\rangle, \\
\hat{\cal H}_{\rm eff}^{\dagger}|\psi_{L}^{}\rangle&=&E^{*}|\psi_{L}^{}\rangle.
\end{eqnarray}

For any two-band Hamiltonian, such as our Hamiltonian in Eq.~\eqref{eq:H_kx_ky}, it can be written as $\hat{\cal H}_{\rm eff}=d_{+}\sigma_{+} + d_{-}\sigma_{-} + d_z\sigma_z=[(d_{+} + d_{-})/2]\sigma_{x} + [i(d_{+} - d_{-})/2]\sigma_{y} + d_{z}\sigma_{z}$ with $d_{\pm}=d_{x} \mp id_{y}$ and $\sigma_{\pm}=(\sigma_{x} \pm i\sigma_{y})/2$, such that we can perform the following parametrization on the Bloch sphere~\cite{qin2024light}:
\begin{eqnarray}
d_{x}&=&d\sin\theta\cos\phi=d\sqrt{1 - D^2}\cos\phi,\\
d_{y}&=&d\sin\theta\sin\phi=d\sqrt{1 - D^2}\sin\phi,\\
d_{z}&=&d\cos\theta=dD,\\
d&=&\sqrt{d_{x}^{2} + d_{y}^{2} + d_{z}^{2}},\\
D&=&\frac{d_{z}}{\sqrt{d_{x}^{2}+d_{y}^{2}+d_{z}^{2}}},\\
R&=&\sqrt{\frac{d_{+}}{d_{-}}}=\sqrt{\frac{d_{x} - id_{y}}{d_{x} + id_{y}}}=\sqrt{\frac{e^{-i\phi}}{e^{i\phi}}}=e^{-i\phi},
\end{eqnarray} where we use $d_{x} + id_{y}=d_{\perp}(\cos\phi + i\sin\phi)=d_{\perp}e^{i\phi}$ and $d_{x} - id_{y}=d_{\perp}(\cos\phi - i\sin\phi)=d_{\perp}e^{-i\phi}$ with $d_{\perp}=\sqrt{d_{x}^{2}+d_{y}^{2}}$.
Hence, our Hamiltonian can be expressed as
\begin{eqnarray}
\hat{\cal H}_{\rm eff}&=&d_x\sigma_x + d_y\sigma_y + d_z\sigma_z
=\begin{pmatrix}
d_{z} & d_{x} - id_{y}  \\
d_{x} + id_{y} & - d_{z}
\end{pmatrix}
=\begin{pmatrix}
dD & d\sqrt{1 - D^2}(\cos\phi - i\sin\phi) \\
d\sqrt{1 - D^2}(\cos\phi + i\sin\phi) & - dD
\end{pmatrix} \nonumber\\
&=&d\begin{pmatrix}
D & \sqrt{1 - D^2}e^{-i\phi} \\
\sqrt{1 - D^2}e^{i\phi} & - D
\end{pmatrix}
=d\begin{pmatrix}
D & \sqrt{1 - D^2}R \\
\sqrt{1 - D^2}R^{-1} & - D
\end{pmatrix}.
\end{eqnarray}

We have the left and right eigenstates as
\begin{eqnarray}
&&|\psi_{R}^{(\pm)}\rangle=\frac{1}{\sqrt{2}}\begin{pmatrix}
\pm R\sqrt{1\pm D} \\
\sqrt{1\mp D}
\end{pmatrix}, \\
&&|\psi_{L}^{(\pm)}\rangle=\frac{1}{\sqrt{2}}\begin{pmatrix}
\pm \sqrt{1\pm D^{*}}/R^{*} \\
\sqrt{1\mp D^{*}}
\end{pmatrix},\\
&&\langle\psi_{R}^{(\pm)}|=\frac{1}{\sqrt{2}}\begin{pmatrix}
\pm R^{*}\sqrt{1\pm D^{*}} &
\sqrt{1\mp D^{*}}
\end{pmatrix}, \\
&&\langle\psi_{L}^{(\pm)}|=\frac{1}{\sqrt{2}}\begin{pmatrix}
\pm \sqrt{1\pm D}/R &
\sqrt{1\mp D}
\end{pmatrix}.
\end{eqnarray}
Here, there are two choices for each bra and ket, and in principle, we can compute four possible inequivalent Berry curvatures, although all four can be shown to integrate to the same Chern number~\cite{shen2018topological}. We choose the biorthogonal option with $\langle\psi_{L}^{(\pm)}|$ bra and $|\psi_{R}^{(\pm)}\rangle$ ket, which is the most physical in terms of the probabilistic interpretation of quantum mechanics~\cite{brody2013biorthogonal}. Doing so, we obtain the biorthogonal Berry curvature as
\begin{eqnarray}
\bar{\Omega}_{xy}^{(\pm)}&=&i\left( \langle\partial_{k_x}\psi_{L}^{(\pm)}|\partial_{k_y}\psi_{R}^{(\pm)}\rangle - \langle\partial_{k_y}\psi_{L}^{(\pm)}|\partial_{k_x}\psi_{R}^{(\pm)}\rangle \right) \nonumber\\
&=&\frac{i}{2}\left[ \begin{pmatrix}
\partial_{k_x}(\pm\sqrt{1\!\pm\! D}/R) &
\partial_{k_x}\sqrt{1\!\mp\! D}
\end{pmatrix}\begin{pmatrix}
\partial_{k_y}( \pm R\sqrt{1\!\pm\! D}) \\
\partial_{k_y}\sqrt{1\!\mp\! D}
\end{pmatrix}  \!-\! \begin{pmatrix}
\partial_{k_y}(\pm\sqrt{1\!\pm\! D}/R) &
\partial_{k_y}\sqrt{1\!\mp\! D}
\end{pmatrix}\begin{pmatrix}
\partial_{k_x}( \pm R\sqrt{1\!\pm\! D}) \\
\partial_{k_x}\sqrt{1\mp D}
\end{pmatrix}\right] \nonumber\\
&=&\frac{i}{2}\left[
\partial_{k_x}(\sqrt{1\!\pm\! D}/R)\partial_{k_y}( R\sqrt{1\!\pm\!D}) \!-\!
\partial_{k_y}(\sqrt{1\!\pm\! D}/R)\partial_{k_x}( R\sqrt{1\!\pm\!D}) \right] \nonumber\\
&=&\frac{i}{2}\left[
\left(\frac{1}{2R\sqrt{1\!\pm\! D}}\partial_{k_x}(\pm D) - \frac{\sqrt{1\!\pm\! D}}{R^2}\partial_{k_x}R \right)\left( \frac{R}{2\sqrt{1\!\pm\!D}}\partial_{k_y}(\pm D) + \sqrt{1\!\pm\!D}\partial_{k_y}R \right) \right.\nonumber\\
&&~\left. \!-\!
\left(\frac{1}{2R\sqrt{1\!\pm\! D}}\partial_{k_y}(\pm D) - \frac{\sqrt{1\!\pm\! D}}{R^2}\partial_{k_y}R \right)\left( \frac{R}{2\sqrt{1\!\pm\!D}}\partial_{k_x}(\pm D) + \sqrt{1\!\pm\!D}\partial_{k_x}R \right) \right] \nonumber\\
&=&\pm\frac{i}{2R}\left[(\partial_{k_x}D)(\partial_{k_y}R)  -  (\partial_{k_y}D)(\partial_{k_x}R) \right]
=\pm\frac{i}{2}\left[(\partial_{k_x}D)(\partial_{k_y}\ln R)  -  (\partial_{k_y}D)(\partial_{k_x}\ln R) \right],
\end{eqnarray} where
\begin{eqnarray}
\partial_{k_{y}}D(\bar{k}_{x},k_{y})=\frac{d[D(\bar{k}_{x},k_{y})]}{dk_{y}}=\frac{\partial[D(\bar{k}_{x},k_{y})]}{\partial k_{y}}+i\frac{\partial[D(\bar{k}_{x},k_{y})]}{\partial \bar{k}_{x}}\frac{d\kappa_{x}(k_{y})}{dk_{y}}.
\end{eqnarray}
Finally, we can get the analytical expression for the biorthogonal Berry curvature as
\begin{equation}
\bar{\Omega}_{xy}^{(\pm)}\!=\!\frac{\mp2t_{y}\Omega_{0}\Omega_{r}\cos\phi_{0}\cos(k_{x}\!+\!i\kappa_{x})\sin k_{y}}{\{[-i\gamma\!+\!2t_{y}\cos k_{y} \!+\! 2t_{x}\cos(k_{x}\!+\!i\kappa_{x})]^{2} \!+\! 4\Omega_{0}^{2}\cos^{2}\phi_{0}\sin^{2}(k_{x}\!+\!i\kappa_{x}) \!+\! [\Omega_{r}\!+\!2\Omega_{0}\sin\phi_{0}\sin(k_{x}\!+\!i\kappa_{x})]^{2}\}^{3/2}}.\label{eq:BC_S1}
\end{equation}
In our cylindrical geometry under $x$ OBCs, non-Hermitian pumping occurs in the $x$ direction, and the complex flux $\kappa_{x}(k_{y})$ contributes to the derivative of $D(\bar{k}_{x},k_{y})$ and ${\rm In}R(\bar{k}_{x},k_{y})$.

If we define the wave numbers as $k_{x}\rightarrow k_{x}a_{x}$ and $k_{y}\rightarrow k_{y}a_{y}$, we can rewrite Eq.~\eqref{eq:BC_S1} as
\begin{equation}
\bar{\Omega}_{xy}^{(\pm)}\!=\!\frac{\mp2a_{x}a_{y}t_{y}\Omega_{0}\Omega_{r}\cos\phi_{0}\cos(k_{x}a_{x}\!+\!i\kappa_{x})\sin (k_{y}a_{y})}{\{[-i\gamma\!+\!2t_{y}\cos (k_{y}a_{y}) \!+\! 2t_{x}\cos(k_{x}a_{x}\!+\!i\kappa_{x})]^{2} \!+\! 4\Omega_{0}^{2}\cos^{2}\phi_{0}\sin^{2}(k_{x}a_{x}\!+\!i\kappa_{x}) \!+\! [\Omega_{r}\!+\!2\Omega_{0}\sin\phi_{0}\sin(k_{x}a_{x}\!+\!i\kappa_{x})]^{2}\}^{3/2}},\label{eq:BC_S2}
\end{equation} where we use
\begin{eqnarray}
\partial_{k_{x}}D(\bar{k}_{x}a_{x},k_{y}a_{y})&=&a_{x}\partial_{(k_{x}a_{x})}D(\bar{k}_{x}a_{x},k_{y}a_{x}),\\
\partial_{k_{y}}D(\bar{k}_{x}a_{x},k_{y}a_{y})&=&a_{y}\partial_{(k_{y}a_{y})}D(\bar{k}_{x}a_{x},k_{y}a_{y}),\\
\partial_{k_{x}}\ln R(\bar{k}_{x}a_{x},k_{y}a_{y})&=&a_{x}\partial_{(k_{x}a_{x})}\ln R(\bar{k}_{x}a_{x},k_{y}a_{x}),\\
\partial_{k_{y}}\ln R(\bar{k}_{x}a_{x},k_{y}a_{y})&=&a_{y}\partial_{(k_{y}a_{y})}\ln R(\bar{k}_{x}a_{x},k_{y}a_{y}).
\end{eqnarray}

\subsection{Robustness of the semiclassical trajectories}\label{Supp_2.3}

\begin{figure}[h]
\centering
\includegraphics[width=0.6\linewidth]{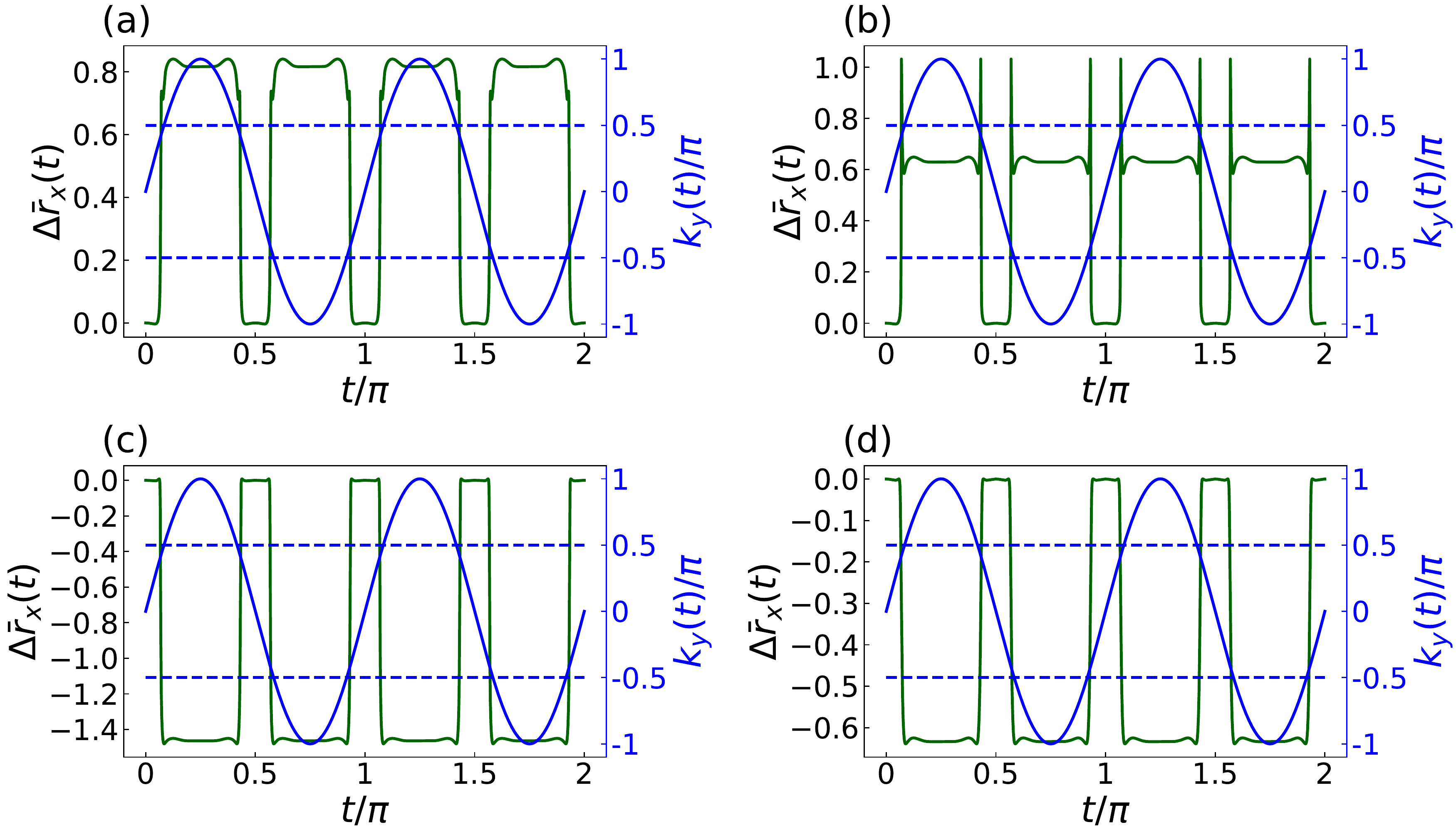}
\caption{The $x$ displacement in Eq.~\eqref{supprdif} (solid green lines). The kinked responses appear when the wave packet momentum $k_{y}(t)=\frac{1}{\hbar}\int_{0}^{t}F_{0}\cos(\omega t^{\prime})dt^{\prime}$ (blue curves) approaches $\pm\pi/2$ under the external fields $F_{x}=12$ and $F_{0}=2\pi$ with $\omega=2$, as marked by the black lines. The parameters are $\phi_{0}=\pi/4$, $t_{x}=\Omega_{0}=2$, $\Omega_{r}=2$, $t_{y}=6$, and $\gamma=2.0,2.5,3.0,3.5$ for (a)-(d). The abrupt jumps occur precisely when the wave packet momentum $k_{y}(t)=\frac{1}{\hbar}\int_{0}^{t}F_{0}\cos(\omega t^{\prime})dt^{\prime}$ (blue curve) approaches $\pm \pi/2$, as marked by the blue dashed lines. Without loss of generality, these parameters are in arbitrary units (arb. units).}\label{fig:suppkink}
\end{figure}

\begin{figure}[h]
\centering
\includegraphics[width=0.6\linewidth]{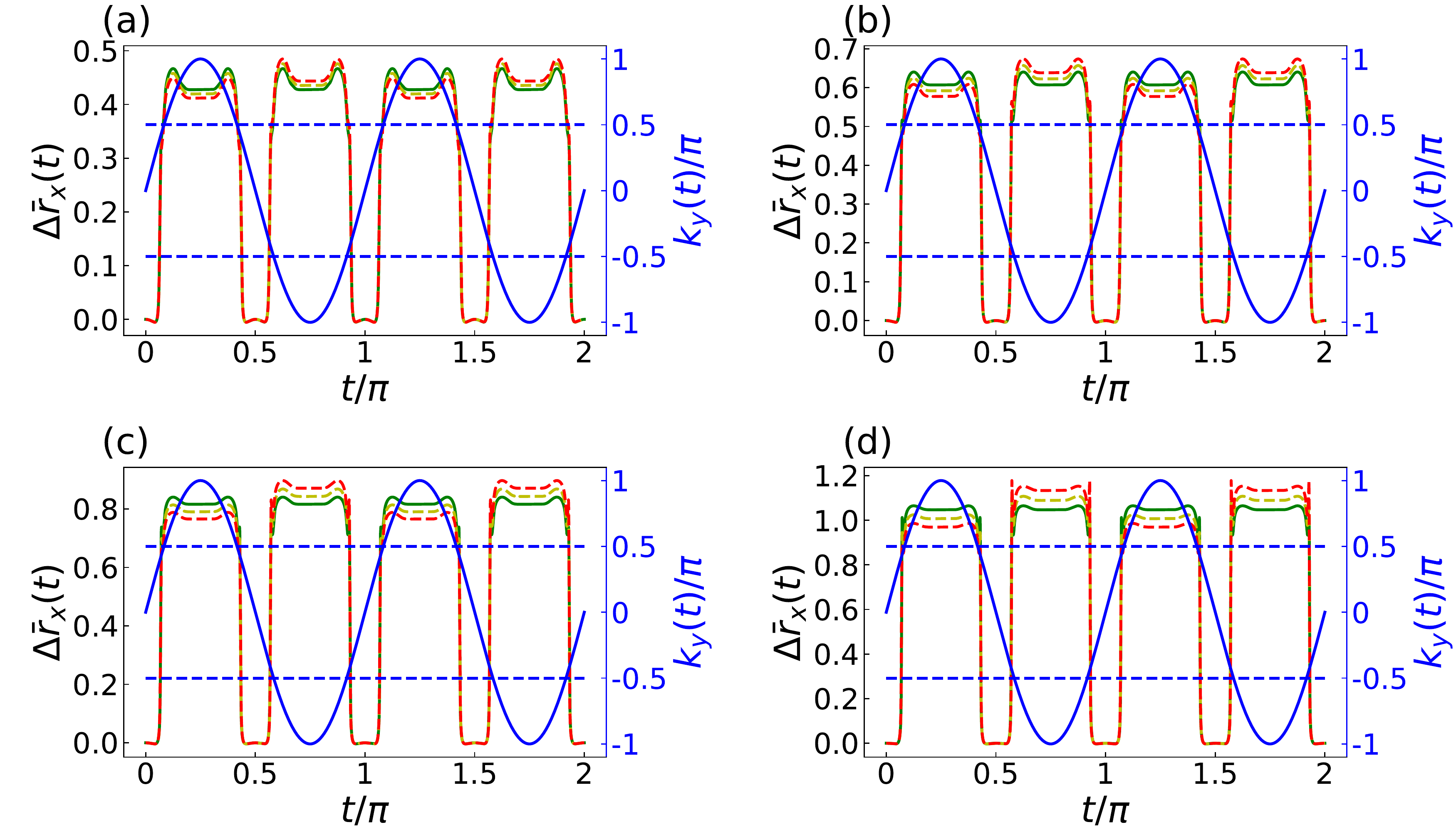}
\caption{The $x$ displacement in Eq.~\eqref{supprdif} (solid green lines) and the perturbative results under $\delta k_{y}=0.01$ (yellow dashed curves) and $0.02$ (red dashed curves) shifts in the initial wave packet. Manifestly, the strongly kinked response (largely abrupt jumps) remains robust. The model is Eq.~\eqref{eq:H_kappa}, with parameters $\phi_{0}=\pi/4$, $t_{x}=\Omega_{0}=2$, $\Omega_{r}=2$, $\gamma=2$, and $t_{y}=4,5,6,7$ for (a)-(d). The external fields are $F_{x}=12$ and $F_{0}=2\pi$ with $\omega=2$. The abrupt jumps occur precisely when the wave packet momentum $k_{y}(t)=\frac{1}{\hbar}\int_{0}^{t}F_{0}\cos(\omega t^{\prime})dt^{\prime}$ (blue curve) approaches $\pm \pi/2$, as marked by the black lines. Without loss of generality, these parameters are in arbitrary units (arb. units).}
\label{fig:suppsemidy}
\end{figure}

Here, we show that the semiclassical response kinks remain very prominent even if the initial wave packet exhibits uncertainty in its momentum distribution. As discussed in the main text, the $x$ displacements of two initially stationary wave packets driven by forces ${\bf F}_{\pm}(t)=F_{0}\cos(\omega t){\bf e}_{y}\pm F_{x}{\bf e}_{x}$  are  given as follows~\cite{longhi2009bloch,longhi2015bloch,xu2017weyl}:
\begin{eqnarray}
\Delta\overline{\mathbf{r}}_{x}(t)&=&\int_{0}^{t}\left[\overline{\mathbf{v}}_{x}(t^{\prime})|_{{\bf F}_{+}}+\overline{\mathbf{v}}_{x}(t^{\prime})|_{{\bf F}_{-}}\right]dt^{\prime} \nonumber\\
&=&-\frac{1}{\hbar}\int_{0}^{t} \left[\frac{\partial [{\rm Re}E(\overline{\mathbf{k}}_{+}(t^{\prime}))]}{\partial k_{x}}+\frac{\partial [{\rm Re}E(\overline{\mathbf{k}}_{-}(t^{\prime}))]}{\partial k_{x}} \right]dt^{\prime}
- \frac{1}{\hbar}\!\int_{0}^{t}[\mathbf{F}_{\pm}(t^{\prime})\!\times\!\mathbf{e}_{z}] \operatorname{Re}\!\left[\bar{\Omega}_{x y}\left(\overline{\mathbf{k}}_{-}\left(t^{\prime}\right)\right)\!+\!\bar{\Omega}_{x y}\left(\overline{\mathbf{k}}_{+}\left(t^{\prime}\right)\right)\right]\! d t^{\prime} \nonumber\\
&=&-\frac{1}{\hbar}\!\int_{0}^{t}F_y(t) \operatorname{Re}\!\left[\bar{\Omega}_{x y}\left(\overline{\mathbf{k}}_{-}\left(t^{\prime}\right)\right)\!+\!\bar{\Omega}_{x y}\left(\overline{\mathbf{k}}_{+}\left(t^{\prime}\right)\right)\right]\! dt^{\prime},\label{supprdif}
\end{eqnarray}
where $\mathbf{v}_{x}$ is the velocity along $x$ direction, $\bar{\Omega}_{xy}({\bf k})$ is the Berry curvature of model Eq.~\eqref{eq:H_kappa}, $\bar{\bf k}_{\pm}(t)={\bf k}_{\pm}(t) + i\kappa_{x}({\bf k}_{\pm}(t))$, and $E(\overline{\mathbf{k}}_{\pm}(t^{\prime}))$ denotes the lower band of model Eq.~\eqref{eq:H_kappa}. Here, the even energy dispersion of the band allows all non-Berry curvature contributions to cancel, leaving only the response purely from the Berry curvature. Thus, Berry curvature discontinuities from singular points of $\kappa_{x}(k_y)$ lead to unique responses in the form of kinked $x$ displacement, as elaborated in the main text. Such kinked responses are manifested as the abrupt jumps (dark green) in Fig.~\ref{fig:suppkink} as the $k_y$ momentum approaches $\pm\pi/2$.

To test the robustness of such jumps or kinks, we slightly perturb the initial momentum distribution of the wave packet from zero as shown in Fig.~\ref{fig:suppsemidy}. Importantly, they are shown to be robust: under the initial $k_{y}$ shifts $\delta k_{y}=0.01$ and $\delta k_{y}=0.02$ in the wave packet, we observe a broadened distribution of $\Delta\overline{\mathbf{r}}_{x}$ (light green). But importantly, the abrupt jump is still very much protected, even though the time when the momentum reaches the Berry curvature discontinuity varies a little.

\subsection{Hermitian case for the energy spectra and Berry curvature}\label{Supp_2.4}

\begin{figure}[h]
\centering
\includegraphics[width=0.5\linewidth]{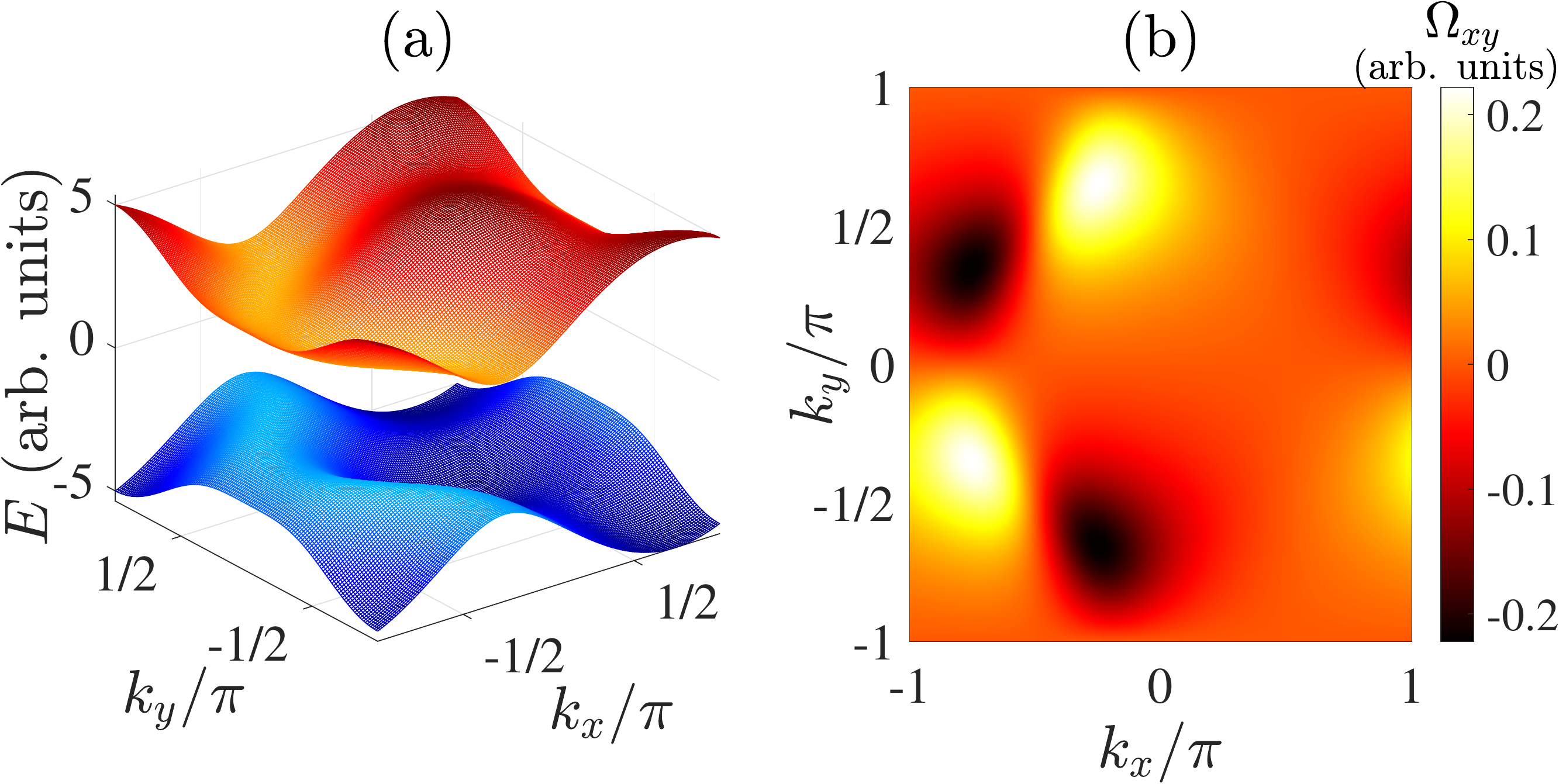}
\caption{(a) Energy spectra (red for the upper band and blue for the lower band) of our cold-atom Hamiltonian ${\cal H}$ in Hermitian case $\gamma=0$ under PBC along $x$ direction and PBC along $y$ direction. (b) Berry curvature $\Omega_{xy}$ of ${\cal H}$ in Hermitian case $\gamma=0$. Parameters for (a) and (b) are $\phi_{0}=\pi/4$, $t_{x}=\Omega_{0}=1$, $t_{y}=t_{x}$, $\Omega_{r}=3t_{x}$, and $\gamma=0$. Without loss of generality, these parameters are in arbitrary units (arb. units).}\label{fig:Hermitian}
\end{figure}

From Fig.~\ref{fig:Hermitian}, it is found that there are no discontinuities in the energy spectrum and Berry curvature. Compared with the non-Hermitian case, we can find that the discontinuities of the energy spectrum and Berry curvature comes from non-Hermitian pumping. So that our kinked linear response comes from the non-Hermiticity.

\twocolumngrid
\bibliography{references_Berry}
\end{document}